\documentclass[preprint2]{proto}

\usepackage{times}
\usepackage{natbib}

\newcommand{\refs}{\par\noindent\hangindent=1pc\hangafter=1}
\newcommand{\avir}      {\alpha_{\rm vir}}
\newcommand{\tkh}        {t_{\rm KH}}
\newcommand{\tsf}        {t_{*f}}

\begin{document}

\title{\textbf{\LARGE The Formation of Massive Stars}}

\author {\textbf{\large Henrik Beuther}}
\affil{\small\em Max-Planck-Institute for Astronomy, Heidelberg}

\author {\textbf{\large Edward B.~Churchwell}}
\affil{\small\em University of Wisconsin, Madison}

\author {\textbf{\large Christopher F.~McKee}}
\affil{\small\em University of California, Berkeley}

\author {\textbf{\large Jonathan C.~Tan}}
\affil{\small\em University of Florida, Gainesville}

\begin{abstract}



\baselineskip = 11pt
\leftskip = 0.65in
\rightskip = 0.65in



\parindent=1pc

{\small ~

  Massive stars have a profound influence on the Universe, but their
  formation remains poorly understood. We review the current status of
  observational and theoretical research in this field, describing the
  various stages of an evolutionary sequence that begins with cold,
  massive gas cores and ends with the dispersal and ionization of gas
  by the newly-formed star. The physical processes in massive star
  formation are described and related to their observational
  manifestations. Feedback processes and the relation of massive stars
  to star cluster formation are also discussed. We identify key
  observational and theoretical questions that future studies should
  address.
  \\~\\~\\~}


\end{abstract}  

\section{Introduction}

Massive star formation has drawn considerable interest for several
decades, but the last 10 years have witnessed a strong acceleration of
theoretical and observational research in this field. One of the major
conceptual problems in massive star formation arises from the
radiation pressure massive stars exert on the surrounding dust and gas
core (e.g., {\em Kahn}, 1974; {\em Wolfire and Cassinelli}, 1987; {\em
  Jijina and Adams}, 1996; {\em Yorke and Sonnhalter}, 2002; {\em
  Krumholz et al.}, 2005b). In principle, this radiation pressure
could be strong enough to stop further accretion, which would imply
that the standard theory of low-mass star formation had to be adapted
to account for the formation of massive stars.  Two primary approaches
have been followed to overcome these problems: the first and more
straightforward approach is to modify the standard theory
quantitatively rather than qualitatively. Theories have been proposed
that invoke varying dust properties (e.g., {\em Wolfire and
  Cassinelli}, 1987), increasing accretion rates in turbulent cloud
cores of the order $10^{-4}-10^{-3}$\,M$_{\odot}$\,yr$^{-1}$ compared
to $\sim 10^{-6}$\,M$_{\odot}$\,yr$^{-1}$ for low-mass star formation
(e.g., {\em Norberg and Maeder}, 2000; {\em McKee and Tan}, 2003),
accretion via disks (e.g., {\em Jijina and Adams}, 1996; {\em Yorke
  and Sonnhalter}, 2002), accretion through the evolving hypercompact
H{\sc ii} region ({\em Keto}, 2003; {\em Keto and Wood}, 2006), the
escape of radiation through wind-blown cavities ({\em Krumholz et
  al.}, 2005a) or radiatively driven Rayleigh-Taylor instabilities
({\em Krumholz et al.}, 2005b). These variations to the standard
picture of low-mass star formation suggest that massive stars can form
within an accretion-based picture of star formation. Contrary to this,
a paradigm change for the formation of massive stars has been proposed
based on the observational fact that massive stars always form at the
dense centers of stellar clusters: the coalescence scenario. In this
scenario, the protostellar and stellar densities of a forming
massive cluster are high enough ($\sim 10^8$\,pc$^{-3}$) that
protostars undergo physical collisions and merge, thereby avoiding the
effects of radiation pressure ({\em Bonnell et al.}, 1998; {\em Bally
  and Zinnecker}, 2005).  Variants of the coalescence model that
operate at lower stellar densities have been proposed by {\em Stahler
  at al.}  (2000) and by {\em Bonnell and Bate} (2005). A less
dramatic approach suggests that the bulk of the stellar mass is
accreted via competitive accretion in a clustered environment ({\em
  Bonnell et al.}, 2004).  This does not necessarily require the
coalescence of protostars, but the mass accretion rates of the massive
cluster members would be directly linked to the number of their
stellar companions, implying a causal relationship between the cluster
formation process and the formation of higher-mass stars therein.

We propose an evolutionary scenario for massive star formation, and
then discuss the various stages in more detail. Following {\em
  Williams et al.}  (2000), we use the term {\it clumps} for
condensations associated with cluster formation, and the term {\it
  cores} for molecular condensations that form single or
gravitationally bound multiple massive protostars.  The evolutionary
sequence we propose for high-mass star-forming cores is:\\ \indent
High-Mass Starless Cores (HMSCs)\\ $\rightarrow$ High-Mass Cores
harboring accreting Low/Intermediate- \indent Mass Protostar(s)
destined to become a high-mass
star(s) \\ $\rightarrow$ High-Mass Protostellar Objects (HMPOs)\\
$\rightarrow$ Final Stars.\\ The term HMPO is used here
in a literal sense, i.e., accreting high-mass protostars. Hence, the
HMPO group consists of protostars $>$8\,M$_{\odot}$, which early on
have not necessarily formed a detectable Hot Molecular Core (HMC)
and/or Hypercompact H{\sc ii} region (HCH{\sc ii}s, size $<0.01$\,pc).
HMCs and HCH{\sc ii}s might coexist simultaneously. Ultracompact H{\sc
  ii} regions (UCH{\sc ii}s, size $<0.1$\,pc) are a transition group:
some of them may still harbor accreting protostars (hence are at the
end of the HMPO stage), but many have likely already ceased accretion
(hence are part of the Final-Star class). High-mass
stars can be on the main sequence while they are deeply embedded and
actively accreting as well as after they cease accreting and become
Final Stars. The class of High-Mass Cores harboring
accreting Low/Intermediate-Mass Protostars has not been well studied
yet, but there has to be a stage between the HMSCs and the HMPOs,
consisting of high-mass cores with embedded low/intermediate-mass
objects.  On the cluster/clump scale the
proposed evolutionary sequence is:\\ \indent Massive Starless Clumps\\
$\rightarrow$ Protoclusters\\ $\rightarrow$ Stellar Clusters.\\
By definition, Massive Starless Clumps can harbor only HMSCs (and
low-mass starless cores), whereas Protoclusters in principle can
harbor all sorts of smaller-scale entities (low- and intermediate-mass
protostars, HMPOs, HMCs, HCH{\sc ii}s, UCH{\sc ii}s and even HMSCs).

This review discusses the evolutionary stages and their associated
physical processes (\S\ref{initial}, \ref{hmpo}, \ref{feedback},
\ref{conclusion}), feedback processes (\S\ref{feedback}), and cluster
formation (\S \ref{cluster}), always from an observational {\it and}
theoretical perspective. We restrict ourselves to present day massive
star formation in a typical Galactic environment. Primordial star
formation, lower metallicities or different dust properties may change
this picture (e.g., {\em Bromm and Loeb}, 2004; {\em Draine}, 2003).
The direct comparison of the theoretical predictions with the
observational evidences and indications shows the potentials and
limitations of our current understanding of high-mass star formation.
We also refer to the IAU227 Proceedings dedicated to Massive
Star Birth ({\em Cesaroni et al.}, 2005b).

\section{Initial conditions of massive star/cluster formation}
\label{initial}

\subsection{Observational results}
\label{irdc}

The largest structures within our Galaxy are Giant Molecular Clouds
(GMCs) with sizes from $\sim$20 to $\sim$100\,pc and masses between
$\sim$10$^4$ to $\sim$10$^6$\,M$_{\odot}$. The physical properties
have been discussed in many reviews (e.g., {\em McKee}, 1999; {\em
  Evans}, 1999), and we summarize only the most important
characteristics.  A multi-transition survey of GMCs in our Galaxy
shows that the average local density derived from an LVG analysis is
$\rho_{\rm H}\sim 4\times 10^3-1.2\times 10^4$\,cm$^{-3}$ and the
temperature is $\sim 10-15$\,K (e.g., {\em Sanders et al.}, 1993),
giving a typical Bonnor-Ebert mass $\sim$ 2\,M$_\odot$.  The
volume-averaged densities in GMCs are $\rho_{\rm H}\sim$ 50 to
100\,cm$^{-3}$; these are substantially less than the local density
values, indicating that the molecular gas is highly clumped.  Velocity
dispersions of 2-3\,km\,s$^{-1}$ indicate highly supersonic internal
motions given that the typical sound speed is $\sim$0.2\,km\,s$^{-1}$.
These motions are largely due to turbulence (e.g., {\em MacLow and
  Klessen}, 2004; {\em Elmegreen and Scalo}, 2004).  Measured magnetic
field strengths are of the order a few $10\,\mu$G (e.g., {\em
  Crutcher}, 1999; {\em Bourke et al.}, 2001). Depending on the
size-scales and average densities, magnetic critical masses can range
from $\sim 5\times 10^5$\,M$_{\odot}$ to a few solar masses,
corresponding to GMCs and low-mass star-forming regions, respectively
({\em McKee}, 1999). Thus, although the rather low Jeans masses
indicate gravitationally bound and likely unstable entities within
GMCs, turbulence and magnetic stresses appear to be strong enough to
support the GMCs against complete collapse on large scales.

Most important for any star formation activity, GMCs show
sub-structures on all spatial scales. They contain dense gas
clumps that are easily identifiable in the (sub)mm continuum and
high-density molecular line tracers (e.g., {\em Plume et al.}, 1992;
{\em Bronfman et al.}, 1996; {\em Beuther et al.}, 2002a; {\em Mueller
  et al.}, 2002; {\em Faundez et al.}, 2004; {\em Beltr{\'a}n et al.},
2006). Peak densities in such dense clumps can easily reach
$10^6$\,cm$^{-3}$, and the massive dense clumps we are interested in
typically have masses between a few 100 and a few 1000\,M$_{\odot}$
(e.g., {\em Beuther et al.}, 2002a; {\em Williams et al.}, 2004; {\em
  Faundez et al.}, 2004). Massive dense clumps are the main locations
where high-mass star formation is taking place. We shall concentrate
on the physical properties and evolutionary stages of clumps of dense
molecular gas and dust.

\begin{figure}[htb] \begin{center}
\includegraphics[width=8.4cm]{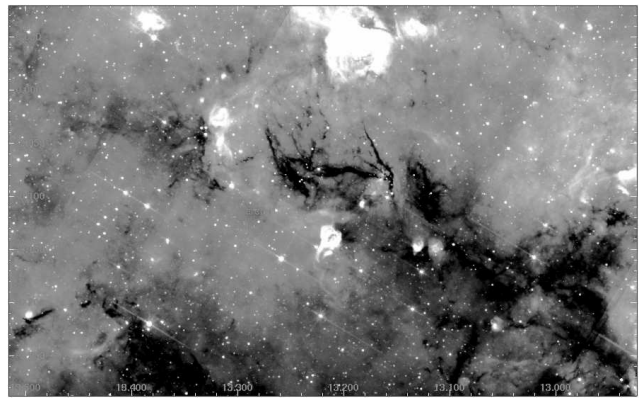}\\ \end{center}
\caption{Example image of IRDCs observed against the Galactic
background with the SPITZER GLIMPSE survey in the 8\,$\mu$m band ({\em
Benjamin et al.}, 2003).}  \label{glimpse} \end{figure}

Most observational high-mass star formation research in the last
decade has focused on HMPOs and UCH{\sc ii} regions. These objects
have mid-infrared emission from hot dust and thus already contain an
embedded massive protostellar source. Earlier evolutionary stages at
the onset of massive star formation were observationally largely
inaccessible because no telescope existed to identify these objects.
The most basic observational characteristics of the earliest stages of
massive star formation, prior to the formation of any embedded heating
source, should be that they are strong cold dust and gas emitters at
(sub)mm wavelengths, and weak or non-detections in the mid-infrared
because they have not yet heated a warm dust cocoon. The advent of the
mid-infrared Space Observatories ISO and MSX permitted for the first
time identifications of large samples of potential (Massive) Starless
Clumps, the Infrared Dark Clouds (IRDCs, e.g., {\em Egan et al.},
1998; {\em Bacmann et al.}, 2000; {\em Carey et al.}, 2000). Fig.
\ref{glimpse} shows a series of IRDCs as seen with SPITZER/GLIMPSE.
Various groups work currently on massive IRDCs, but so far not much
has been published. Some initial ideas about observational quantities
at the initial stages of massive cloud collapse were discussed by {\em
Evans et al.} (2002). {\em Garay et al.} (2004) presented early mm
observations of a sample of 4 sources, other recent statistical
identifications and studies of potential HMSCs or regions at the onset
of massive star formation can be found in {\em Hill et al.}~(2005),
{\em Klein et al.}~(2005), {\em Sridharan et al.}~(2005), and {\em
Beltr{\'a}n et al.}~(2006).

These massive dense clumps have masses between a few 100 and a few
1000\,M$_{\odot}$, sizes of the order 0.25-0.5\,pc, mean densities of
$10^5$\,cm$^{-3}$, and temperatures of order 16\,K. While the masses
and densities are typical for high-mass star-forming regions, the
temperatures derived, for example, from NH$_3$ observations (around
16\,K, {\em Sridharan et al.}, 2005) are lower than those toward young
HMPOs and UCH{\sc ii} regions (usually $\geq 22$\,K, e.g., {\em
  Churchwell et al.}, 1990; {\em Sridharan et al.}, 2002).
Furthermore, the measured NH$_3$ line-widths from the IRDCs are narrow
as well; {\em Sridharan et al.}~(2005) found mean values of
1.6\,km\,s$^{-1}$ whereas HMPOs and UCH{\sc ii}s have mean values of
2.1 and 3.0\,km\,s$^{-1}$, respectively ({\em Churchwell et al.},
1990; {\em Sridharan et al.}, 2002).  The narrow line-widths and low
temperatures support the idea that IRDCs represent an earlier
evolutionary stage than HMPOs and UCH{\sc ii} regions, with less
internal turbulence. Although subject to large uncertainties, a
comparison of the virial masses calculated from NH$_3$ data with the
gas masses estimated from 1.2\,mm continuum emission indicates that
most candidate HMSCs are virially bound and prone to potential
collapse and star formation ({\em Sridharan et al.}, 2005).

The IRDCs are not a well defined class, but are expected to harbor
various evolutionary stages, from genuine HMSCs via High-Mass Cores
harboring accreting Low/Intermediate-Mass Protostars to the youngest
HMPOs. While the first stage provides good targets to study the initial
conditions of massive star formation prior to cloud collapse, the
other stages are important to understand the early evolution of
massive star-forming clumps. For example, the source HMSC18223-3 is
probably in such an early accretion stage: Correlating
high-spatial-resolution mm observations from the Plateau de Bure
Interferometer with SPITZER mid-infrared observations, {\em Beuther et
  al.}~(2005c) studied a massive dust and gas core with no
protostellar mid-infrared counterpart in the GLIMPSE data. While this
could also indicate a genuine HMSC, they found relatively high
temperatures ($\sim$33\,K from NH$_3$(1,1) and (2,2)), an
increasing N$_2$H$^+$(1--0) line-width from the core edge to the core
center, and so called "green fuzzy" mid-infrared emission at the edge
of the core in the IRAC data at 4.5\,$\mu$m, indicative of molecular
outflow emission. The outflow scenario is supported by strong
non-Gaussian line-wing emission in CO(2--1) and CS(2--1). These
observational features are interpreted as circumstantial evidence for
early star formation activity at the onset of massive star formation.
Similar results toward selected IRDCs have recently been reported by,
e.g., {\em Rathborne et al.}~(2005), {\em Ormel et al.}~(2005), {\em
  Birkmann et al.}~(2006) and {\em Pillai et al.}~(2006).
Interestingly, none of these IRDC case studies has revealed a true
HMSC yet. However, with the given low number of such studies, we
cannot infer whether this is solely a selection effect or whether
HMSCs are genuinely rare.

Recent large-scale mm-continuum mapping of Cygnus-X revealed
approximately the same number of infrared-quiet sources compared with
infrared-bright HMPO-like regions ({\em Motte et al.}, 2005). However,
none of these infrared-quiet sources appears to be a genuine HMSC, and
hence they could be part of the class of High-Mass Cores harboring
accreting Low/Intermediate-Mass Protostars. A few studies report
global infall on large spatial scales (e.g., {\em Rudolph et al.},
1990; {\em Williams and Garland}, 2002; {\em Peretto et al.}, 2006;
{\em Motte et al.}, 2005), suggesting that massive clumps could form
from lower-density regions collapsing during the early star
formation process.

The earliest stages of massive star formation, specifically massive
IRDCs, have received increased attention over the past few years, but
some properties like the magnetic field have so far not been studied
at all. Since this class of objects is observationally rather new, we
are expecting many exciting results in the coming years.

\subsection{Theory}

A central fact about GMCs is that they are turbulent ({\em Larson},
1981). The level of turbulence can be characterized by the virial
parameter $\avir\equiv 5\sigma^2 R/GM$, where $\sigma$ is the rms 1D
velocity dispersion, $R$ the mean cloud radius, and $M$ the cloud
mass; $\avir$ is proportional to the ratio of kinetic to gravitational
energy ({\em Myers and Goodman}, 1988; {\em Bertoldi and McKee}, 1992).
The large-scale surveys of GMCs by {\em Dame et al.} (1986) and {\em
Solomon et al.} (1987) give $\avir\simeq 1.3-1.4$ ({\em McKee and Tan},
2003), whereas regions of active low-mass star formation have
$\avir\simeq 0.9$ (e.g., {\em Onishi et al.}, 1996). For regions of
massive star formation, {\em Yonekura et al.} (2005) find $0.5\la
\avir\la 1.4$.

The great advance in our understanding of the dynamics of GMCs in the
past decade has come from simulations of the turbulence in GMCs (e.g.,
{\em Elmegreen and Scalo}, 2004; {\em Scalo and Elmegreen}, 2004; {\em
  MacLow and Klessen}, 2004).  One of the primary results of these
studies is that turbulence decays in less than a crossing time.  A
corollary of this result is that it is difficult to transmit turbulent
motions for more than a wavelength.  These results raise a major
question: since most of the sources of interstellar turbulence are
intermittent in both space and time, how is it possible to maintain
the high observed levels of turbulence in the face of such strong
damping? Simulations without driven turbulence, such as those used to
establish the theory of competitive accretion (e.g., {\em Bonnell et
  al.}, 2001a; see \S 3.2), reach virial parameters $\alpha_{\rm
  vir}\ll 1$, far less than observed.

The results of these turbulence simulations have led to two competing
approaches to the modeling of GMCs and the gravitationally bound
structures (clumps) within them: as quasi-equilibrium structures or as
transient objects.  The first approach builds on the classical
analysis of {\em Spitzer}, (1978) and utilizes the steady-state virial
theorem ({\em Chieze}, 1987; {\em Elmegreen}, 1989; {\em Bertoldi and
McKee}, 1992; {\em McKee}, 1999). This model naturally explains why
GMCs and the bound clumps within them have virial parameters of order
unity (provided the magnetic field has a strength comparable to that
observed) and why their mean column densities in the Galaxy are $\sim
10^{22}$ H cm$^{-2}$.  In order to account for the ubiquity of the
turbulence, such models must assume that: (1) the turbulence actually
decays more slowly than in the simulations, perhaps due to an
imbalanced MHD cascade ({\em Cho and Lazarian}, 2003); (2) the energy
cascades into the GMC or clump from larger scales; and/or (3) energy
injection from star formation maintains the observed level of
turbulence ({\em Norman and Silk}, 1980; {\em McKee}, 1989; {\em
Matzner}, 2002). Recent simulations by {\em Li and Nakamura} (in
prep.)  are consistent with the suggestion that protostellar energy
injection can indeed lead to virial motions in star-forming clumps
(see \S 4.2).  In the alternative view, the clouds are transient and
the observed turbulence is associated with their formation ({\em
Ballesteros-Paredes et al.}, 1999; {\em Elmegreen}, 2000; {\em
Hartmann et al.}, 2001; {\em Clark et al.}, 2005; {\em Heitsch et
al.}, 2005). {\em Bonnell et al.}~(2006) propose that the observed
velocity dispersion in molecular clouds could be due to clumpy
molecular gas passing through galactic spiral shocks.  While these
theories naturally account for the observed turbulence, they do not
explain why GMCs have virial parameters of order unity, nor do they
explain why clouds that by chance live longer than average do not have
very low levels of turbulence.  Quasi-equilibrium models predict that
star formation will occur over a longer period of time than do
transient cloud models. How these predictions compare with
observations of high-mass star formation regions will be discussed in
\S \ref{cluster_thy} below.

\section{High-Mass Protostellar Objects}
\label{hmpo}

\subsection{Observational results}

\underline{General properties:} The most studied objects in massive
star formation research are HMPOs and UCH{\sc ii} regions.  This is
partly because the IRAS all sky survey permitted detection and
identification of a large number of such sources from which
statistical studies could be undertaken (e.g., {\em Wood and
  Churchwell}, 1989a; {\em Plume et al.}, 1992; {\em Kurtz et al.}
1994; {\em Shepherd and Churchwell}, 1996; {\em Molinari et al.}, 1996; {\em
  Sridharan et al.}, 2002; {\em Beltr{\'a}n et al.}, 2006). The main
observational difference between young HMPOs/HMCs and UCH{\sc ii}
regions is that the former are weak or non-detections in the cm-regime
due to undetectable free-free emission (for a UCH{\sc ii} discussion
see, e.g., {\em Churchwell}, 2002, and the chapter by {\em Hoare et al.}).
Although in our classification typical HMCs with their high
temperatures and complex chemistry are a subset of HMPOs, we expect
that every young HMPO must already have heated a small central gas
core to high temperatures, and it is likely that sensitive
high-spatial resolution observations will reveal small HMC-type
structures toward all HMPOs. This is reminiscent of the so-called Hot
Corinos found recently in some low-mass star-forming cores (see the
chapter by {\em Ceccarelli et al.}).

Many surveys have been conducted in the last decade characterizing the
physical properties of massive star-forming regions containing HMPOs
(e.g., {\em Plume et al.}, 1997; {\em Molinari et al.}, 1998; {\em
  Sridharan et al.}, 2002; {\em Beuther et al.}, 2002a; {\em Mueller
  et al.}, 2002; {\em Shirley et al.}, 2003; {\em Walsh et al.}, 2003;
{\em Williams et al.}, 2004; {\em Faundez et al.}, 2004; {\em Zhang et
  al.}, 2005; {\em Hill et al.}, 2005; {\em Klein et al.}, 2005; {\em
  Beltr{\'a}n et al.}, 2006). While the masses and sizes are of the same
order as for the IRDCs (a few 100 to a few 1000\,M$_{\odot}$ and of
the order 0.25-0.5\,pc, \S\ref{initial}), mean densities can exceed
$10^6$\,cm$^{-3}$, and mean surface densities, although with a
considerable spread, are reported around 1\,g\,cm$^{-2}$ (for a
compilation see {\em McKee and Tan}, 2003).  In contrast to earlier
claims that the density distributions $\rho\propto r^{-k_{\rho}}$ of massive
star-forming clumps may have power-law indices $k_{\rho}$ around $1.0$,
several studies derived density distributions with mean power-law
indices $k_{\rho}$ around $1.5$ ({\em Beuther et al.}, 2002a; {\em Mueller et
  al.}, 2002; {\em Hatchell and van der Tak}, 2003; {\em Williams et
  al.}, 2004), consistent with density distributions observed toward
regions of low-mass star formation (e.g., {\em Motte and Andr{\'e}}, 2001).
However, one has to bear in mind that these high-mass studies analyzed
the density distributions of the gas on cluster-scales whereas the
low-mass investigations trace scales of individual or multiple
protostars. Mean temperatures ($\sim$22\,K, derived from NH$_3$
observations) and NH$_3$(1,1) line widths
($\sim$2.1\,km\,s$^{-1}$) are also larger for HMPOs than for IRDCs.

Furthermore, HMPOs are often associated with H$_2$O and Class {\sc ii}
CH$_3$OH maser emission (e.g., {\em Walsh et al.} 1998; {\em Kylafis
  and Pavlakis}, 1999; {\em Beuther et al.}, 2002c; {\em Codella et
  al.}, 2004; {\em Pestalozzi et al.}, 2005; {\em Ellingsen}, 2006).
While the community agrees that both maser types are useful signposts
of massive star formation (H$_2$O masers are also found in low-mass
outflows), there is no general agreement what these phenomena actually
trace in massive star-forming regions. Observations indicate that both
species are found either in molecular outflows (e.g., {\em De Buizer},
2003; {\em Codella et al.}, 2004) or in potential massive accretion
disks (e.g., {\em Norris et al.}, 1998; {\em Torrelles et al.}, 1998).
In a few cases, such as very high spatial resolution VLBI studies, it
has been possible to distinguish between an origin in a disk and an
outflow (e.g., {\em Torrelles et al.}, 2003; {\em Pestalozzi et al.},
2004; {\em Goddi et al.}, 2005), but, in general, it is mostly not
possible to distinguish between the two possibilities.

One of the most studied properties of HMPOs are the massive
molecular outflows found to be associated essentially with all stages
of early massive star formation. For a discussion of this phenomenon
and its implications see \S\ref{feedback}.

\underline{Massive disks:} Disks are an essential property of the
accretion-based formation of high-mass stars. The chapter by
{\em Cesaroni et al.} provides a detailed discussion about observations
and modeling of disks around massive protostellar objects. We simply
summarize that massive, Keplerian disks have been observed around
early-B-type stars, the best known example being IRAS\,20126+4102
({\em Cesaroni et al.}, 1997, 1999, 2005a; {\em Zhang et al.}, 1998).
Venturing further to higher-mass sources, several studies found
rotating structures perpendicular to molecular outflows, indicative of
an inner accretion disk (e.g., {\em Zhang et al.}, 2002; {\em Sandell
  et al.}, 2003; {\em Beltr{\'a}n et al.}, 2004; {\em Beuther et al.},
2005b).  However, these structures are not necessarily Keplerian and
could be larger-scale toroids, rotating around the central forming O-B
cluster as suggested by {\em Cesaroni} (2005).  Recently {\em van der
  Tak and Menten} (2005) conclude from 43\,GHz continuum observations
that massive star formation at least up to $10^5$\,L$_{\odot}$
proceeds through accretion with associated collimated molecular
outflows. A detailed theoretical and observational understanding of
massive accretion disks is one of the important issues for future
high-mass star formation studies.

\underline{SEDs:} Spectral energy distributions (SEDs) have often been
used to classify low-mass star-forming regions and to infer their
physical properties (e.g., {\em Lada and Wilking}, 1984; {\em Andr{\'e} et
al.}, 1993). In massive star formation, deriving SEDs for
individual high-mass protostellar sources proves to be more
complicated. Problems arise because of varying spatial resolution with
frequency, varying telescope sensitivity, and disk orientation to the
line of sight. While we can resolve massive cluster-forming regions in
the (sub)mm regime (e.g., {\em Cesaroni et al.}, 1999; {\em Shepherd
et al.}, 2003; {\em Beuther and Schilke}, 2004) and at cm
wavelengths (e.g., {\em Wood and Churchwell}, 1989b; {\em Kurtz et
al.}, 1994; {\em Gaume et al.}, 1995, {\em De Pree et al.}, 2000),
near-/mid and far-infrared wavelength data for individual
sub-sources are difficult to obtain. The earliest evolutionary
stages are generally so deeply embedded that they are undetectable
at near-infrared wavelengths, and until recently this was also a
severe problem at mid-infrared wavelength (although a few notable
exceptions exist, e.g., {\em De Buizer et al.}, 2002; {\em Linz et
al.}, 2005).  The advent of the SPITZER Space Telescope now allows
deep imaging of such regions and will likely reveal many
objects. However, even with SPITZER the spatial resolution in the
far-infrared regime, where the SEDs at early evolutionary stages
peak, is usually not good enough ($16''$ pixels) to spatially
resolve the massive star-forming clusters.  The only statistically
relevant data at far-infrared wavelengths so far stem from the IRAS
satellite, which had a spatial resolution of approximately
$100''$. Although the IRAS data have proven very useful in
identifying regions of massive star formation (e.g., {\em Wood and
Churchwell}, 1989a; {\em Molinari et al.}, 1996; {\em Sridharan et
al.}, 2002; {\em Beltr{\'a}n et al.}, 2006), they just give the
fluxes integrated over the whole star-forming cluster and thus
hardly constrain the emission of individual cluster members.
Nevertheless, the IRAS data were regularly employed to estimate the
integrated luminosity of young massive star-forming regions by
two-component grey-body fits (e.g., {\em Hunter et al.}, 2000; {\em
Sridharan et al.}, 2002), and to set additional constraints on the
density distributions of the regions (e.g., {\em Hatchell and van der
Tak}, 2003). SED modeling allows one to infer the characteristics of
protostars in massive star-forming regions ({\em Osorio et al.},
1999).  {\em Chakrabarti and McKee} (2005) showed that the far-IR SEDs
of protostars embedded in homogeneous, spherical envelopes are
characterized by the density profile in the envelope and by two
dimensionless parameters, the light-to-mass ratio, $L/M$, and the
surface density of the envelope, $\Sigma=M/(\pi R^2)$. If these
parameters are determined from the SED and if one knows the distance,
then it is possible to infer both the mass and accretion rate of the
protostar ({\em McKee and Chakrabarti}, 2005).  {\em Whitney et al.}
(2005) and {\em De Buizer et al.} (2005) have determined the effects
of disks on the SEDs.  Recent 3-D modeling by {\em Indebetouw et al.}
(2006) shows how sensitive the SEDs, especially below $100\,\mu$m, are
to the clumpy structure of the regions and to the observed line of
sights. For example, they are able to fit the entire sample of UCH{\sc
  ii} regions studied by {\em Faison et al.}  (1998) with the same
clumpy model because the varying line of sights produce very different
SEDs. Hence, SEDs alone do not provide sufficient information to infer
the properties of clumpy sources, and it will be essential to obtain
additional information by mapping these sources with powerful
observatories such as the SMA, CARMA in 2006, ALMA at the end of the
decade and JWST in the next decade.

\underline{Chemistry:} Young massive star-forming regions, and
specifically HMCs, exhibit a rich chemistry from simple two-atom
molecules to large organic carbon chains (e.g., {\em Blake et al.},
1987; {\em Schilke et al.}, 1997, 2001). While these single-dish
observations were not capable of resolving the chemical differences
within the regions, interferometric studies toward a few sources have
revealed the spatial complexity of the chemistry in HMCs (e.g., {\em
  Blake et al.}, 1996; {\em Wyrowski et al.}, 1999; {\em Beuther et
  al.}, 2005a). Here, we present studies toward W3(H$_2$O)/OH and
Orion-KL as prominent chemical show-cases.

{\it W3(H$_2$O):} The Hot Core region W3(H$_2$O) $6''$ east of the
UCH{\sc ii} region W3(OH) exhibits an H$_2$O maser outflow and a
synchrotron jet ({\em Alcolea et al.}, 1993; {\em Wilner et al.},
1999). Follow-up observations with the Plateau de Bure Interferometer
(PdBI) reveal dust emission associated with the synchrotron jet source
and a large diversity of molecular line emission between the UCH{\sc
  ii} region W3(OH) and the Hot Core W3(H$_2$O) ({\em Wyrowski et
  al.}, 1997, 1999).  Nitrogen-bearing molecules are observed only
toward W3(H$_2$O), whereas oxygen-bearing species are detected from
both regions. Based on HNCO observations, {\em Wyrowski et al.} (1999)
estimate gas temperatures toward W3(H$_2$O) of $\sim$200\,K, clearly
confirming the hot core nature of the source. The differences in
oxygen- and nitrogen-bearing species are manifestations of chemical
evolution due to different ages of the sources.

\begin{figure}[htb] \begin{center}
\includegraphics[angle=-90,width=8.4cm]{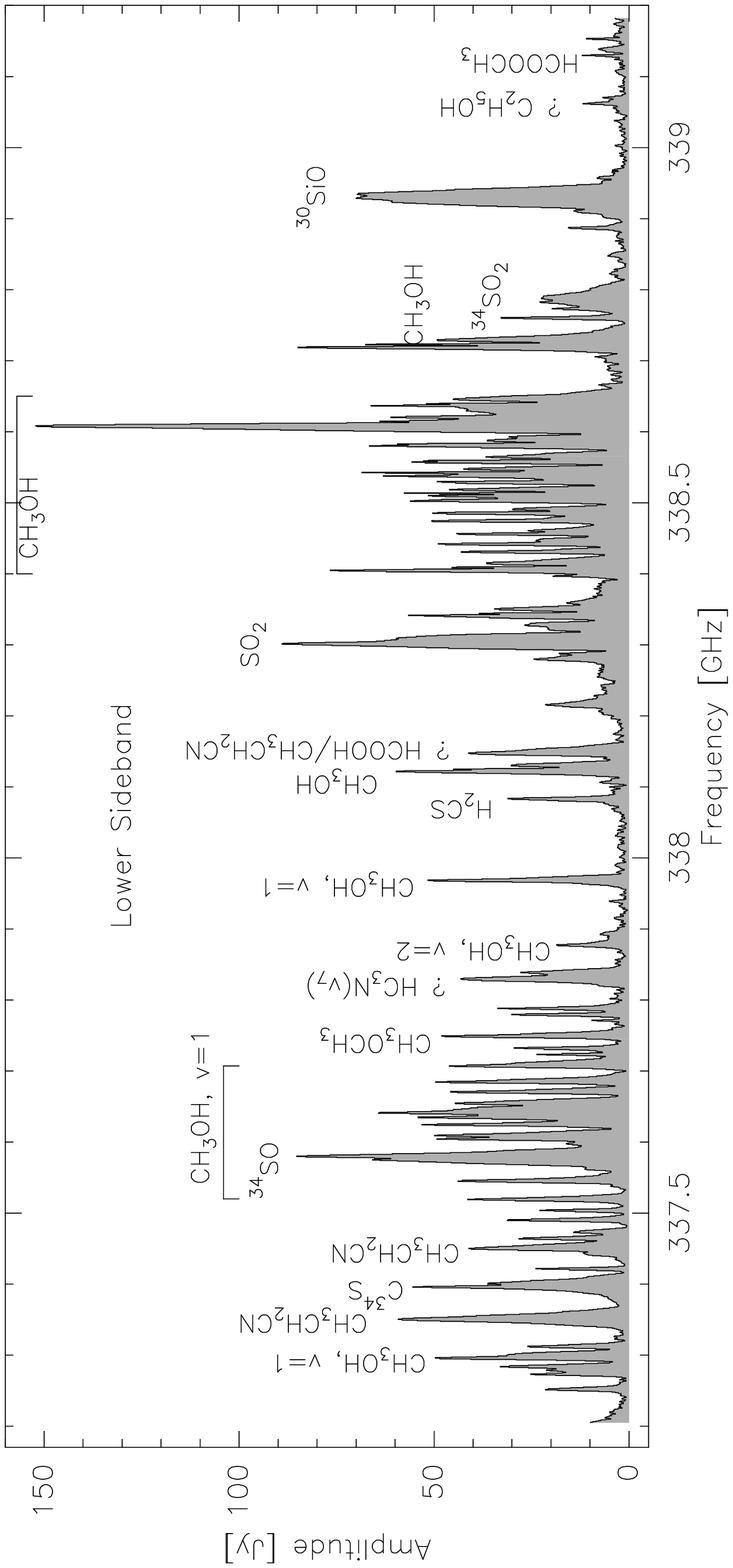}\\
\includegraphics[angle=-90,width=8.4cm]{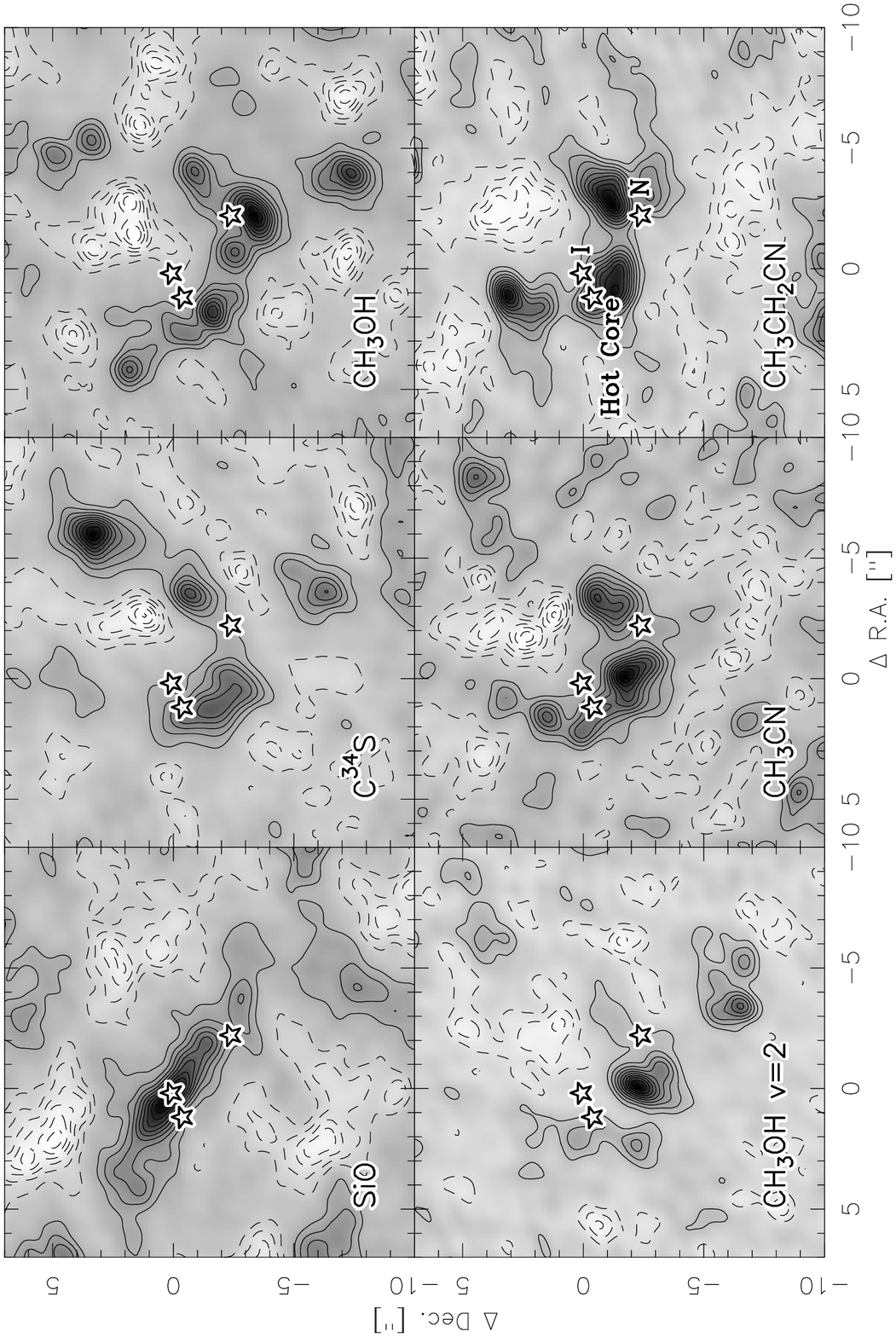} \end{center}
\caption{Submillimeter Array spectral line observations in the
  850\,$\mu$m band toward Orion-KL from {\em Beuther et al.}  (2005a).
  The top panel shows a vector-averaged spectrum in the uv-domain on a
  baseline of 21\,m. The bottom panel presents representative images
  from various molecular species. Full contours show positive
  emission, dashed contours negative features due to missing short
  spacings and thus inadequate cleaning. The stars mark the locations
  of source {\it I}, the Hot Core peak position and source {\it n}
  (see bottom-right panel).}  \label{orion}
\end{figure}

{\it Orion-KL:} One of the early targets of the recently completed
Submillimeter Array (SMA) was the prototypical HMC Orion-KL. {\em
  Beuther et al.} (2004, 2005a, 2006) observed the region in the
865\,$\mu$m and 440\,$\mu$m windows and resolved the submm continuum
and molecular line emission at $1''$ resolution. The continuum maps
resolved the enigmatic source {\it I} from the hot molecular core,
detected source {\it n} for the first time shortward of 7\,mm and
furthermore isolated a new protostellar source SMA1, emitting strong
line emission.  The observed 4\,GHz bandpass in the 865\,$\mu$m band
revealed more than 145 lines from various molecular species with
considerable spatial structure. Fig. \ref{orion} shows an SMA
example spectrum and representative line images. SiO emission is
observed from the collimated north-east south-west outflow and the
more extended north-west south-east outflow. Typical hot core
molecules like CH$_3$CN and CH$_3$CH$_2$CN follow the hot core
morphology known from other molecules and lower frequency observations
(e.g., {\em Wright et al.}, 1996; {\em Blake et al.}, 1996; {\em
  Wilson et al.}, 2000). In contrast to this, oxygen-bearing molecules
like CH$_3$OH or HCOOCH$_3$ are weaker toward the hot molecular core,
but they show strong emission features a few arcseconds to the
south-west, associated with the so-called compact ridge. Many
molecules, in particular sulphur-bearing species like C$^{34}$S or
SO$_2$, show additional emission further to the north-east, associated
with IrC6.

Although existing chemical models predict the evolution and production
paths of various molecules (e.g., {\em Charnley}, 1997; {\em van
Dishoeck and Blake}, 1998; {\em Doty et al.}, 2002; {\em Nomura and
Millar}, 2004; {\em Viti et al.}, 2004), we are certainly not at the
stage where they can reliably predict the chemical
structure of HMCs.  Considering the complexity of the closest region
of massive star formation, Orion-KL, it is essential to get a deeper
understanding of the basic chemical and physical processes, because
otherwise the confidence in studies of regions at larger distances
is greatly diminished. On the positive side, a better knowledge of
the chemical details may allow us to use molecular line observations
as chemical clocks for (massive) star-forming regions.

\subsection{Theory}
\label{hmpo_theory}

The critical difference between low- and high-mass star formation is
that low-mass stars form in a time $\tsf$ short compared to the
Kelvin-Helmholtz time $\tkh$, whereas high-mass stars generally have
$\tkh\la\tsf$ ({\em Kahn} 1974). As a result, low-mass stars undergo
extensive pre-main sequence evolution after accretion has finished,
whereas the highest mass stars can accrete a significant amount of
mass while on the main sequence. The feedback associated with the
intense radiation produced by high-mass stars will be considered in \S
4; here we ask whether high-mass star formation differs significantly
from low-mass star formation. At the time of the last Protostars and
Planets conference, {\em Stahler et al.} (2000) argued that it does.

The conventional view remains that high-mass star formation is a
scaled up version of low-mass star formation, with an accretion rate
$\dot m_*\simeq c^3/G$, where the effective sound speed $c$ includes
the effects of thermal gas pressure, magnetic pressure, and turbulence
({\em Stahler et al.}, 1980). {\em Wolfire and Cassinelli} (1987)
found that accretion rates of order $10^{-3}\;M_\odot$~yr$^{-1}$ are
needed to overcome the effects of radiation pressure, and attributed
this to the high values of $c$ in high-mass star forming regions. By
modeling the SEDs of high-mass protostars, {\em Osorio et al.} (1999)
inferred that high-mass stars form in somewhat less than $10^5$
yr. They favored a logatropic model, in which the ambient density
varies as $r^{-1}$ away from the protostar.  {\em McKee and Tan}
(2002, 2003) critiqued logatropic models and developed the Turbulent
Core Model, in which massive stars form from gravitationally bound
cores supported by turbulence and magnetic fields.  They argued that
on scales large compared to the thermal Jeans mass, the density and
pressure distributions in turbulent, gravitationally bound cores and
clumps should be scale free and vary as powers of the radius (e.g.,
$\rho\propto r^{-k_{\rho}}$).  As a result, the core and the
star-forming clump in which it is embedded are polytropes, with
$P\propto \rho^{{\gamma_p}}$.  The gravitational collapse of a
polytrope that is initially in approximate equilibrium results in an
accretion rate $\dot m_*\propto m_*^q$, with
$q={3(1-\gamma_p)/(4-3\gamma_p)}$ ({\em McLaughlin and Pudritz},
1997).  Isothermal cores have $q=0$, whereas logatropes
($\gamma_p\rightarrow 0$) have $q=\frac 34$.  (It should be noted that
the numerical simulations of {\em Yorke and Sonnhalter}, 2002,
generally have $q<0$ due to feedback effects. This simulation differs
from the turbulent core model in that the initial conditions were
non-turbulent and the restriction to two dimensions overemphasizes
feedback effects--\S 4.2.) Regions of high-mass star formation have
surface densities $\Sigma\sim 1$\,g\,cm$^{-2}$ ({\em Plume et al.,}
1997), corresponding to visual extinctions $A_V \sim 200$~mag.  {\em
  McKee and Tan} (2003) showed that the typical accretion rate and the
corresponding time to form a star of mass $m_{*f}$ in such regions are
\setlength{\arraycolsep}{0.5mm} \begin{eqnarray*} \dot m_*&\simeq&
  0.5\times 10^{-3}\left(\frac{m_{*f}}{30 M_\odot} \right)^{3/4}
  \Sigma_{\rm cl}^{3/4}\left(\frac{m_*}{m_{*f}}\right)
  ^{0.5}~~\frac{M_\odot}{\mbox{yr}},\\ \tsf&\simeq &1.3\times 10^5
  \left(\frac{m_{*f}}{30\;M_\odot}\right) ^{1/4}\Sigma_{\rm
    cl}^{-3/4}~~~\mbox{yr}, \end{eqnarray*} where $\Sigma_{\rm cl}$ is
the surface density of the several thousand $M_\odot$ clump in which
the star is forming and where they adopted $k_{\rho}=\frac 32$ as a
typical value for the density power law in a core. The radius of the
core out of which the star forms is $0.06(m_{*f}/30M_\odot)^{1/2}
\Sigma_{\rm cl}^{1/2}$~pc.  Observed star clusters in the Galaxy have
surface densities comparable to those of high-mass star forming
regions, with values ranging from about 0.2 g cm$^{-2}$ in the Orion
Nebula Cluster to about 4 g cm$^{-2}$ in the Arches Cluster ({\em
  McKee and Tan}, 2003).  This work has been criticized on two
grounds: First, it approximates the large-scale macro-turbulence in
the cores and clumps as a local pressure (micro-turbulence), which is
equivalent to ignoring the surface terms in the virial equation (e.g.,
{\em Mac Low} 2004).  This approximation is valid provided the cores
and clumps live for a number of free-fall times, so that they are in
quasi-equilibrium. Evidence that clumps are quasi-equilibrium
structures will be discussed in \S 5.2; being smaller, cores are
likely to experience greater fluctuations, so the quasi-equilibrium
approximation is probably less accurate for them.  Second, the
turbulent core model assumes that most of the mass in the core that is
not ejected by outflows will go into a single massive star (or
binary).  {\em Dobbs et al.} (2005) investigated this assumption by
simulating the collapse of a high-mass core similar to that considered
by {\em McKee and Tan} (2003). In the isothermal case, {\em Dobbs et
  al.}  found that the core fragmented into many pieces, which is
inconsistent with the formation of a massive star. With a more
realistic equation of state, however, only a few fragments formed, and
when the heating due to the central protostar is included, even less
fragmentation occurs ({\em Krumholz et al.} 2005b). Furthermore, the
level of turbulence in the simulations by {\em Dobbs et al.} (2005) is
significantly less than the observed value.

Variants of the gravitational collapse model in which the accretion
rate accelerates very rapidly have also been considered ($\dot
m_*\propto m_*^q$ with $q>1$, so that $m_*\rightarrow \infty$ in a
finite time in the absence of other effects). However, such models
have accretion rates that can exceed the value $\dot m_*\propto c^3$
that is expected on dynamical grounds.  {\em Behrend and Maeder}
(2001) assumed that the accretion rate onto a protostar is
proportional to the observed mass loss rate in the protostellar
outflow and found that a massive star could form in $\sim 3\times
10^5$ yr. This phenomenological model has $q\simeq 1.5$, the value
adopted in an earlier model by {\em Norberg and Maeder} (2000).
However, it is not at all clear that the accretion rate onto the
protostar is in fact proportional to the observed mass outflow rates.
{\em Keto} (2002, 2003) modeled the growth of massive stars as being
due to Bondi accretion, so that the accretion rate is $\dot m_*\propto
m_*^2$, under his assumption that the ambient medium has a constant
density and temperature.  As {\em Keto} points out, the Bondi
accretion model assumes that the self-gravity of the gas is
negligible. The condition that the mass within the Bondi radius
$Gm_*/c^2$ be much less than the stellar mass can be shown to be
equivalent to requiring $\dot m_*\la c^3/G$; for the value of $c\simeq
0.5$ km s$^{-1}$ considered by {\em Keto}, this restricts the
accretion rate to $\dot m_*\la 3\times 10^{-5} M_\odot$ yr$^{-1}$,
smaller than the values he considers. (One can show that when one
generalizes the Bondi accretion model to approximately include the
self gravity of the gas, the accretion rate is indeed about $c^3/G$ if
the gas is initially in virial equilibrium.) {\em Schmeja and Klessen}
(2004) analyze mass accretion rates in the framework of
gravo-turbulent fragmentation, and they find that the accretion rates
are highly time-variant, with a sharp peak shortly after the formation
of the protostellar core. Furthermore, in their models the peak and
mean accretion rates increase with increasing mass of the final star.

Most models for (proto)stellar structure and evolution do not yet
include the effects of rotation (e.g., {\em Meynet and Maeder} 2005),
which are expected to be relatively large given the recent
accumulation of stellar material from the accretion disk. In models of
gravo-turbulent fragmentation, {\em Jappsen and Klessen} (2004) find
that the angular momentum $j$ correlates with the core mass $M$ like
$j\propto M^{2/3}$. Furthermore, they conclude that the angular
momentum evolution is approximately consistent with the contraction of
initially uniform density spheres undergoing solid body rotation. The
precise amount of stellar angular momentum depends on how the
accretion and outflow from the star-disk interaction region is
modulated by magnetic fields and on the strength of the stellar wind
(e.g., {\em Matt and Pudritz}, 2005). One potentially important effect
is the variation in photospheric temperature from the equatorial to
polar regions, which can enhance the beaming of bolometric and
ionizing luminosity along the polar directions.

\begin{figure*}[htb] 
\includegraphics[width=17cm]{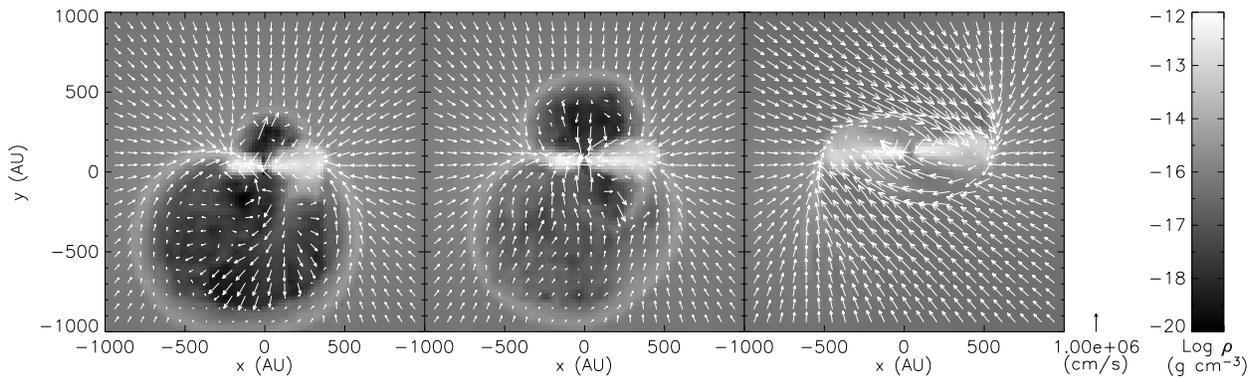} 
\caption{This plot from {\em Krumholz et al.} (2005b) shows 3D
radiation hydrodynamic simulations of the collapse of a massive
core. It is a slice in the XY plane at three different times showing
the initial growth, instability and collapse of a radiation
bubble. The times of the three slices are $1.5\times 10^4$,
$1.65\times 10^4$, and $2.0\times 10^4$ years, and the (proto)stellar
masses are 21.3, 22.4 and 25.7\,M$_{\odot}$. The density is shown in
gray-scale and the velocity as arrows.}  
\label{krumholz}
\end{figure*}

Alternative models for high-mass star formation have been developed by
{\em Bonnell} and collaborators (e.g., {\em Bonnell et al.}, 1998,
2004).  In the competitive accretion model, small stars ($m_*\sim 0.1
M_\odot$) form via gravitational collapse, but then grow by
gravitational accretion of gas that was initially unbound to the
star~-- i.e., by Bondi-Hoyle accretion, with allowance for the
possibility that tidal effects can reduce the accretion radius ({\em
  Bonnell et al.}, 2001a, 2004).  This model naturally results in
segregating high-mass stars toward the center of the cluster, as
observed. Furthermore, it gives a two-power law IMF that is
qualitatively consistent with observation ({\em Bonnell et al.}
2001b).  Simulations by {\em Bonnell et al.} (2004) are consistent
with this model. However, there are two significant difficulties:
First, radiation pressure disrupts Bondi-Hoyle accretion once the
stellar mass exceeds $\sim 10 M_\odot$ ({\em Edgar and Clarke}, 2004),
so it is unlikely that competitive accretion can operate at masses
above this. There is no evidence for a change in the IMF in this mass
range, however, which suggests that competitive accretion does not
determine the IMF at lower masses either.  Second, competitive
accretion is effective only if the virial parameter is much less than
observed: Based on simulations of Bondi-Hoyle accretion in a turbulent
medium, {\em Krumholz et al.} (2005c, 2006) show that protostars of
mass $m_*\sim 0.1 M_\odot$ can accrete more than their initial mass in
a dynamical time only if $\avir \la 0.1(10^3 M_\odot/M_{\rm
  cl})^{1/2}$.  Such low values of $\avir$ do appear in the
simulations, but, as discussed above, not in observed high-mass
star-forming regions, which have masses $M_{\rm cl}$ of hundreds to
thousands of solar masses. Since the expected amount of mass accreted
in a dynamical time is small, {\em Krumholz et al.} conclude that
stars form via gravitational collapse of individual cores
(Fig.~\ref{krumholz}). {\em Bonnell et al.} (this volume) argue that
both difficulties can be ameliorated if the clump in which the massive
stars are forming is undergoing global gravitational collapse. {\em
  Tan et al.} (in prep.) present arguments against such dynamical
collapse in the formation of star clusters.  It is thus very important
to observationally determine the nature of the motions in massive
star-forming clumps: are they dominated by turbulence or by collapse?

The most radical and imaginative model for the formation of high-mass
stars is that they form via stellar collisions during a brief epoch in
which the stellar density reaches $\sim 10^8$ stars pc$^{-3}$ (e.g.,
{\em Bonnell et al.}, 1998; {\em Bonnell and Bate}, 2002), far greater
than observed in any Galactic star cluster (the densest region
reported so far is W3~IRS5 with an approximate stellar density of
$10^6$ stars pc$^{-3}$, {\em Megeath et al.}, 2005). This model also
results in an IMF that is in qualitative agreement with observation,
although it must be borne in mind that the simulations to date have
not included feedback.  In their review, {\em Stahler et al.} (2000)
supported the merger model, emphasizing that gas associated with
protostars could increase the effective collision cross section and
permit merging to occur at lower stellar densities.  More recently,
{\em Bonnell and Bate} (2005) have suggested that binaries in clusters
will evolve to smaller separations due to accretion, resulting in
mergers. However, a key assumption in this model is that there is no
net angular momentum in the accreted gas, which makes sense in the
competitive accretion model but not the gravitational collapse model.
Stellar dynamical calculations by {\em Portegies Zwart et al.} (2004),
which did not include any gas, show that at densities $\ga 10^8$ stars
pc$^{-3}$ it is possible to have runaway stellar mergers at the center
of a star cluster, which they suggest results in the formation of an
intermediate mass black hole. It should be noted that they inferred
that this could have occurred based on the currently observed
properties of the star cluster (although with the assumption that the
tidal radius is greater than 100 times the core radius), not on a
hypothetical ultra-dense state of the cluster.  {\em Bally and
  Zinnecker} (2005) discuss observational approaches to testing the
merger scenario, and suggest that the wide-angle outflow from OMC-1 in
the Orion molecular cloud could be due to a protostellar merger that
released $10^{48}-10^{49}$\,erg.  While it is quite possible that some
stellar mergers occur near the centers of some star clusters, the
hypothesis that stellar mergers are responsible for a significant
fraction of high-mass stars faces several major hurdles: (1) the
hypothesized ultra-dense state would be quite luminous due to the
massive stars, yet has never been observed; (2) the mass loss
hypothesized to be responsible for reducing the cluster density from
$\sim 10^8$ stars pc$^{-3}$ to observed values must be finely tuned in
order to decrease the magnitude of the binding energy by a large
factor; and (3) it is difficult to see how this model could account
for the observations of disks and collimated outflows discussed in \S
3.1 and 4.1.

\section{Feedback processes in massive star formation}
\label{feedback}

\subsection{Observational results}

\underline{Hypercompact H{\sc ii} regions:} Hypercompact H{\sc ii}
regions (HCH{\sc ii}s) are smaller, denser, and brighter than UCH{\sc
ii} regions.  Specifically, they are defined as having diameters less
than 0.01\,pc, consistent with being small photoionized nebulae
produced by O or B stars. None have more ionizing flux than can be
provided by a single O or B star. Their common properties that
distinguish them from UCH{\sc ii} regions are:\\ 1) They are $\ga 10$
times smaller ($\leq 0.01$\,pc) and $\sim$100 times denser than
UCH{\sc ii} regions with emission measures $\geq 10^8$\,pc\,cm$^{-6}$
({\em Kurtz}, 2002, 2005).\\ 2) They have rising radio spectral
indices $\alpha$ (where $S_{\nu} \propto \nu^{\alpha}$) from short cm
to mm wavelengths, and $\alpha$ ranges from $\sim$0.3 to 1.6 with a
typical value of $\sim$1 (e.g., {\em Churchwell}, 2002; {\em Hofner et
al.}, 1996).  They are very faint or not detected at wavelengths
longward of 1\,cm.  The power-law spectra span too large a range in
frequency to be the transition from optically thick to thin emission
in a constant density nebula.\\ 3) In massive star formation regions,
they often appear in tight groups of two or more components ({\em
Sewilo et al.}, 2004), reminiscent of the Trapezium in Orion.\\ 4)
Many but not all HCH{\sc
  ii} regions have unusually broad radio recombination lines (RRLs;
FWHM$\geq$40\,km\,s$^{-1}$). Some have FWHMs$>$100\,km\,s$^{-1}$ ({\em
  Sewilo et al.}, 2004).\\ 5) They are often (always?)  coincident
with strong water masers (e.g., {\em Hofner and Churchwell}, 1996;
{\em Carral et al.}, 1997) and possibly other masers also, but the
latter has not yet been observationally established.\\ What is the
nature of HCH{\sc ii} regions?  Their compactness, multiplicity, range
of luminosities, and coincidence with water masers all argue for
ionization by a single or possibly a binary system of late O or B
star(s) at an age younger than UCH{\sc ii} regions.  Their broad RRLs
indicate highly dynamic internal structures (outflow jets, disk
rotation, expansion, shocks, accretion, etc.) the nature of which is
not yet understood.  The fact that only about half of the HCH{\sc ii}s
that have observed RRLs have broad lines would argue that this phase
is short-lived, perhaps only apparent in the first half of an HCH{\sc
  ii} region's lifetime, provided the accretion rate is larger during
the early stages of the HCH{\sc ii}s. Their radio spectral indices
have implications for the internal density structure, but here also
too little observational information is available to do more than
speculate at this juncture.  The power-law spectra can be produced by
a clumpy nebula ({\em Ignace and Churchwell}, 2004), but this is only
one of several possibilities that needs to be investigated (e.g., {\em
  Keto}, 2003; {\em Tan and McKee}, 2003). It is not clear yet whether
they form after the HMC stage or whether HMCs and HCH{\sc ii}s
coexist.

\underline{Outflows:} Massive molecular outflows are among the most
studied phenomena in massive star formation over the last decade, and
the observations range from statistical studies of large samples at
low spatial resolution to individual case studies at high spatial
resolution. Because (massive) molecular outflows are presented in the
chapter by {\em Arce et al.}, here we only discuss their general
properties and implications for the massive star-forming processes.

Since the early statistical work by {\em Shepherd and Churchwell}
(1996) it is known that massive molecular outflows are an ubiquitous
phenomenon in massive star formation. Early observations indicated
that massive outflows appear less collimated than their low-mass
counterparts, implying potentially different formation scenarios for
the outflows and the massive star-forming processes (e.g., {\em Richer
  et al.}, 2000; {\em Shepherd et al.}, 2003). However, {\em Beuther
  et al.} (2002b) showed that outflows from HMPOs, even if observed
only with single-dish instruments, are consistent with collimated
outflows if one considers the large distances, projection effects and
poor angular resolution carefully (see also {\em Kim and Kurtz},
2006). Interferometric follow-up studies revealed more massive
star-forming regions with molecular outflows consistent with the
collimated outflows known from low-mass star formation (for a
compilation see {\em Beuther and Shepherd}, 2005).  Since collimated
structures are hard to maintain over a few $\times 10^4$ years (the
typical dynamical timescales of molecular outflows) if they are
associated with colliding protostars within the cluster centers, these
outflow observations strongly support the accretion-based formation
scenario in massive star formation.

We note that no highly collimated outflow has been observed for
high-mass star-forming regions exceeding $10^5$\,L$_{\odot}$,
corresponding to approximately 30\,M$_{\odot}$ stars. Therefore, these
data cannot exclude that stars more massive than that may form via
different processes. However, there are other possibilities to explain
the current non-observations of collimated outflows at the
high-luminosity end. An easy explanation would be that these sources
are so exceptionally rare that we simply have not been lucky enough to
detect one.  Alternatively, {\em Beuther and Shepherd} (2005) recently
suggested an evolutionary sequence that explains qualitatively the
present state of observational facts (see also the chapter by {\em
Arce et al.}): To form massive early O-stars via accretion, the
protostellar objects have to go through lower-mass stages as well.
During the early B-star stage, the accreting protostars can drive
collimated outflows as observed. Growing further in mass and
luminosity, they develop HCH{\sc ii} regions in the late O-star stage,
and collimated jets and less collimated winds can coexist, producing
bipolar outflows with a lower degree of collimation.  In this
scenario, it would be intrinsically impossible to ever observe
jet-like outflows from young early O-type protostars. Alternatively,
the effect may be due to greater observational confusion of the
outflows from very luminous sources with those from surrounding
lower-mass protostars, since the more luminous sources tend to be in
richer, more distant clusters. These evolutionary models for massive
molecular outflows have to be tested further against theory and
observations.

\subsection{Theory}

Feedback processes that act against gravitational collapse and
accretion of gas to protostars include radiation pressure (transmitted
via dust grains, and, for sufficiently massive stars, by electron
scattering), thermal pressure of photo-ionized gas, ram pressure from
protostellar winds, and main sequence stellar winds.  These processes
become increasingly important with protostellar mass and may reduce
the efficiency of star formation from a given core.  There is good
evidence for a cutoff in the stellar IMF at around 150~$M_\odot$
(e.g., {\em Weidner and Kroupa}, 2004; {\em Figer}, 2005), but it
remains to be determined whether this is due to feedback processes or
to instabilities in massive stars.

For individual low-mass star formation from a core, bipolar
protostellar outflows, accelerated from the inner accretion disk and
star by rotating magnetic fields, appear to be the dominant feedback
mechanism, probably preventing accretion from polar directions and
also ejecting a fraction, up to a third, of the material accreting
through the disk. This leads to star formation efficiencies from the
core of order 30-50\% ({\em Matzner and McKee}, 2000).

For massive protostars, forming in the same way from a core and
accretion disk, one expects similar MHD-driven outflows to be present,
leading to similar formation efficiencies. In addition, once the
massive protostar has contracted to the main sequence (this can occur
rapidly before accretion has finished), it starts to produce a large
flux of ionizing photons. The resulting HCH{\sc ii} region is likely
to be confined in all but the polar directions by the protostellar
jets ({\em Tan and McKee}, 2003).  This could provide an important
potential observational diagnostic for the physics of protostellar
jets, as they might be illuminated along the axis by the ionizing
radiation.  As the protostellar mass and ionizing flux increase, the
HCH{\sc ii} region can eventually burn its way through the jet and
begin to ionize the disk surface. If the disk is ionized out to a
radius where the escape speed is about equal to the ionized gas sound
speed, then a photo-evaporated flow is set up, reducing accretion to
the star ({\em Hollenbach et al.}, 1994).

Observations indicate that outflows may be less well collimated for
luminosities above about $10^5$\,L$_{\odot}$ (\S 4.1). As discussed
above, {\em Beuther and Shepherd} (2005) have suggested that this is
due to a decrease in the collimation of the protostellar jet with
increasing protostellar luminosity.  A possible mechanism for this is
suggested by the work of {\em Fendt and Cameljic} (2002), who simulated
protostellar jets with a large turbulent diffusivity and found that
the collimation decreases as the diffusivity increases. Applying this
picture to massive outflows, the level of turbulence in the accretion
flow would have to grow as the luminosity of the protostar increases.

The importance of massive molecular outflows in driving turbulence in
molecular clouds is not generally agreed upon. {\em MacLow and
  Klessen} (2004, and references therein) argue that although
molecular outflows are very energetic, they deposit most of their
energy at low densities. Furthermore, since the molecular gas motions
show increasing power all the way up to the largest cloud complexes,
{\em MacLow and Klessen} (2004) conclude that it would be hard to
fathom how such large scales should be driven by embedded protostars.
Contrary to this, on the relatively small scales of the clumps, if the
energy of turbulent motions decays with a half-life of one dynamical
time, then protostellar outflows from star formation are able to
maintain turbulence if 50\% of the gas mass forms stars in 20
dynamical times, and 1\% of the resulting outflow energy couples to
the ambient gas ({\em Tan}, 2006). Recently, {\em Quillen et al.}
(2005) reported that their observations toward the low-mass
star-forming region NGC1333 are also consistent with outflow driven
turbulence, and, as remarked in \S 2.2, {\em Li and Nakamura} (in
prep.)  have given theoretical support to this idea. It becomes more
difficult for this mechanism to support turbulence on larger scales in
the GMC involving greater gas masses; on these scales, {\em Matzner}
(2002) has shown that energy injection by H{\sc ii} regions dominates
that by protostellar outflows and can support the observed level of
turbulence.  Alternatives to protostellar driving of the turbulence in
molecular clouds are discussed in \S 2.2.

Radiation pressure on dust grains (well-coupled to the gas at these
densities) is also important for massive protostars. It has been
suggested, in the context of spherical accretion models, that this
leads to an upper limit to the initial mass function ({\em Kahn},
1974; {\em Wolfire and Cassinelli}, 1987). The difficulties faced by
spherical accretion models was a major motivation for the formation
model via stellar collisions ({\em Bonnell et al.}, 1998). However,
massive star formation becomes easier once a disk geometry is allowed
for (e.g., {\em Nakano}, 1989; {\em Jijina and Adams}, 1996). {\em
  Yorke and Sonnhalter} (2002) used 2D axially symmetric simulations
to follow massive star formation from a core collapsing to a disk,
including radiation pressure feedback; accretion stopped at
$43\:M_\odot$ in their most massive core.  They showed the accretion
geometry channeled radiative flux into the polar directions and away
from the disk, terming this the ``flashlight effect''. {\em Krumholz
  et al.} (2005a) found that cavities created by protostellar outflows
increase the flashlight effect, allowing even higher final masses. The
first 3D simulation of massive star formation shows that instabilities
facilitate the escape of radiation and allow the formation of stars
significantly more massive than suggested by 2D calculations
(Fig.~\ref{krumholz}, {\em Krumholz et al.}, 2005b).

The high accretion rates required to form massive stars tend to quench
HCH{\sc ii} regions ({\em Walmsley}, 1995). For spherical accretion,
the density profile in a freely infalling envelope is $\rho\propto
r^{-3/2}$. As a result, the radius of the HCH{\sc ii} region is
$$R_{\rm HCHII}=R_*\exp(S/S_{\rm cr}),$$
where $S$ is the ionizing photon luminosity and 
$$S_{\rm cr}=\frac{\alpha^{(2)}\dot m_*^2}{8\pi \mu_{\rm H}^2Gm_*}=5.6\times 10^{50}\left(\frac{\dot m_{*,-3}^2}{m_{*2}}
\right)~~\mbox{s}^{-1}$$ ({\em Omukai and Inutsuka}, 2002).  Here
$\alpha^{(2)}$ is the recombination coefficient to excited states of
hydrogen, $\dot m_{*,-3}=\dot m_*/(10^{-3}M_\odot$ yr$^{-1}$) and
$m_{*2}=m_*/(100 M_\odot)$; we have replaced $m_p$ in their expression
with $\mu_{\rm H}=2.34\times 10^{-24}$ g, the mass per hydrogen
nucleus. When the radius of the HCH{\sc ii} region is small enough
that the infall velocity exceeds the velocity of an R-critical
ionization front ($2c_i$, where $c_i\simeq 10$ km s$^{-1}$ is the
isothermal sound speed of the ionized gas), the HCH{\sc ii} region is
said to be "trapped" ({\em Keto}, 2002): There is no shock in the
accretion flow and the HCH{\sc ii} region cannot undergo the classical
pressure-driven expansion.  The ionizing photon luminosity $S$
increases rapidly with $m_*$. If the accretion rate depends on stellar
mass such that the critical luminosity $S_{\rm cr}$ is approximately
independent of mass (e.g., the standard {\em McKee and Tan} (2003)
model has $\dot m_*\propto m_*^{1/2}$, so that $S_{\rm cr}=$\,const.),
then the radius of the HCH{\sc ii} region expands as $\exp(S)$ and the
trapped phase is relatively brief.  On the other hand, if $S_{\rm cr}$
increases rapidly with mass, as in the Bondi accretion model ($\dot
m_*\propto m_*^2$), then the expansion is retarded, leading to the
possibility that the trapped phase of the HCH{\sc ii} region could
last for much of the life of the protostar ({\em Keto}, 2003).
However, as outlined in \S \ref{hmpo_theory}, the parameters adopted
by {\em Keto} (2003) are not consistent with the neglect of the
self-gravity of the gas.

The evolution of the HCH{\sc ii} region changes substantially due to
rotation of the infalling gas. The density is significantly lower
above the accretion disk ({\em Ulrich}, 1976), so trapped HCH{\sc ii}
regions will generally expand out to the radius of the accretion disk;
when this is larger than the gravitational radius $R_g=Gm_*/c_i^2$,
then the HCH{\sc ii} region is no longer trapped ({\em McKee and Tan},
in prep.).  {\em Keto and Wood} (2006) have also considered the effects
of disks in massive protostars: They point out that it is possible to
form an ionized accretion disk, and suggest that there is evidence for
this in G10.6-04. 

\section{Formation of star clusters with massive stars}
\label{cluster}

\subsection{Observational results}
\label{cluster_obs}

\underline{IMF:} The formation of the Initial Mass Function (IMF) has
been an important issue in star formation research since the early
work by {\em Salpeter} (1955). For a current summary of IMF studies
see {\em Corbelli et al.} (2005) and references therein, and the
chapter by {\em Bonnell et al.}. One of the questions in the context
of this review is whether the IMF is determined already at the
earliest stages of cluster formation by the initial gravo-turbulent
fragmentation processes of molecular clouds (e.g., {\em Padoan and
Nordlund}, 2002; {\em MacLow and Klessen}, 2004), or whether the IMF
is determined by subsequent processes like competitive accretion or
feedback processes from the underlying star-forming cluster (e.g.,
{\em Bonnell et al.}, 2004; {\em Ballesteros-Paredes et al.}, 2006).
(Sub)mm continuum studies of young low-mass clusters have convincingly
shown that the core mass function at the beginning of low-mass cluster
formation already resembles the stellar IMF ({\em Motte et al.}, 1998;
{\em Johnstone et al.}, 2001; {\em Enoch et al.}, 2006; and the
chapter by {\em Lada et al.}). Because massive star-forming regions
are on average more distant, resolving these clusters is
difficult. However, several single-dish studies of different high-mass
star-forming regions at early evolutionary stages have shown that at
high clump masses, the cumulative mass distributions are consistent
with the the slope of the high-mass stellar IMF ({\em Shirley et al.},
2003; {\em Williams et al.}, 2004; {\em Reid and Wilson}, 2005; {\em
Beltr{\'a}n et al.}, 2006).  Furthermore, the only existing
high-spatial-resolution interferometric study that resolves a massive
star-forming clump into a statistically meaningful number of cores
also finds the core mass distribution to be consistent with the
stellar IMF ({\em Beuther and Schilke}, 2004). Although mm continuum
observations alone are ambiguous whether the observed cores and clumps
are bound or transient structures, the consistently steeper mass
functions observed in mm continuum emission compared with the lower
density tracing CO line studies (e.g., {\em Kramer et al.}, 1998)
suggest that the mm-continuum sources could be bound whereas the CO
sources could be transient. Furthermore, {\em Belloche et al.}  (2001)
report additional observations supporting the interpretation that the
study of {\em Motte et al.} (1998) sampled bound sources.  Combining
these results from massive star formation studies with the previous
investigations in the low-mass regime, the apparent similarity between
the (cumulative) clump and core mass functions and the stellar IMF
supports the idea that the IMF is determined by molecular cloud
structure before star formation is initiated, maybe implicating
gravo-turbulent fragmentation. However, on a cautionary note, one has
to keep in mind that the cumulative mass distributions from
single-dish studies as reported above trace scales of cluster
formation and thus probably refer to cluster-mass distributions rather
than to the IMF. The only way to assess the relationship between the
fragmentation of initial high-mass star-forming clumps and the
resulting IMF is to carry out high-spatial-resolution interferometric
(sub)mm line and continuum studies of a statistically significant
sample of (very) young massive star-forming regions.

\underline{First GLIMPSE results:} The GLIMPSE survey is providing an
entirely new view of the inner Galaxy with a higher resolution and
sensitivity than ever achieved at mid-infrared wavelengths ({\em
Benjamin et al.}, 2003).  This is enabling a host of new research on
massive star formation as well as many other fields of astronomy.
Unfortunately, UCH{\sc ii} regions are generally saturated and too
bright in the IRAC bands to identify the ionizing star(s) and
associated clusters above the glaring diffuse PAH emission found
toward all these objects.  The known HCH{\sc ii} regions are also
bright in the GLIMPSE survey. Bipolar outflows stand out in the
4.5\,$\mu$m band, providing a powerful way to identify many new
outflows in the inner Galaxy. Numerous outflows have been identified
in the survey and a catalog of them is being assembled by the GLIMPSE
team.  The mechanism responsible for this emission is believed to be
line emission from shocked H$_2$ and/or CO bands; the shocks are
produced by outflowing gas ramming into the ambient interstellar
medium. Near-infrared spectroscopic observations of molecular H$_2$ or
CO between 4.1 and 4.7\,$\mu$m are needed to determine which
interpretation is correct.

Within $\sim 45^{\circ}$ of the Galactic Center, hundreds of IRDCs are
apparent in silhouette against the diffuse infrared background
(Fig.\ref{glimpse}).  A catalog of many IRDCs is being prepared for
publication by the GLIMPSE team.  These clouds are optically thick at
8\,$\mu$m, implying visual extinctions $\geq50$\,mag (see \S
\ref{irdc}). The GLIMPSE images are striking in part because of the
large number of bubbles contained in them; there are about 1.5 bubbles
per square degree on average.  A catalog of 329 bubbles has been
identified and a web accessible image archive will accompany the
archive ({\em Churchwell et al.}, in prep.).  It is found that the
bubbles are associated with H{\sc ii} regions and stellar clusters.
Only three are associated with supernova remnants and none with
planetary nebulae or Wolf Rayet stars. About 1/3 of the bubbles appear
to be produced by the stellar winds and radiation pressure from O and
B stars (i.e., massive star-forming regions).  About 2/3 of the
bubbles have small angular diameters (typically only 3-4 arcmin) and
do not coincide with a radio H{\sc ii} region or known cluster; these
are believed to be driven by B4-B9 stars that have strong enough winds
to form a resolved bubble and have enough UV radiation to excite PAH
features, but not enough UV photons to ionize a detectable HII region.

One of the most exciting prospects from the GLIMPSE survey is the
possibility of identifying the entire population of HMPOs and lower
mass protostars from the approximately 50 million stars in the GLIMPSE
archive.  This is now possible with the large archive of radiative
transfer models of protostars calculated for the entire range of
protostellar masses, the full range of suspected accretion rates, disk
masses, and orientations of the accretion disks to the line of sight
({\em Whitney et al.}, in prep.).  Model photospheres for main
sequence stars and red giants are included in the model archive as
well, so it is possible to distinguish reddened main sequence stars
and red giants from protostars and slightly evolved young stellar
objects.  What makes this archive of models powerful, however, is the
model fitter that will fit the best models to observed SEDs of large
numbers of sources, giving: the mass, spectral types, approximate
evolutionary state, and interstellar extinction for the best fit
models to every source ({\em Robitaille et al.,}, in prep.).  This
will provide a powerful alternative to the classical method of
estimating the global star formation rate in the Galaxy based on
measured UV photon luminosities of radio H{\sc ii} regions and an
assumed initial mass function .

\subsection{Theory}
\label{cluster_thy}

In the local universe, massive star and star cluster formation are
intrinsically linked: massive stars almost always appear to form in
clusters ({\em De Wit et al.}, 2005). We have seen that this is a
natural expectation of models of star formation from cores, since the
accretion rates are higher if the core is pressurized by the weight of
a large clump of gas. This scenario predicts that massive stars
tend to form near the center of the clump and that there can be
extensive star formation (mostly of lower-mass stars) from the clump's
gas while massive star formation is ongoing. 

Presently there is no consensus on whether massive stars form
preferentially at the centers of clusters, since, although they are
often observed in central locations, it is possible that they could
have migrated there by dynamical interactions after their formation.
{\em Bonnell and Davies} (1998) found that the mass segregation time
of clusters with mass-independent initial velocity dispersions was
similar to the relaxation time, $t_{\rm relax}\simeq 0.1 N/({\rm ln}
N) t_{\rm dyn}$ for $N$ equal mass stars, i.e., about 14 crossing
timescales for $N=1000$.  The presence of gas should shorten these
timescales ({\em Ostriker}, 1999). To resolve this issue, we need to
measure the cluster formation time: does it take few or many dynamical
times?  {\em Elmegreen} (2000) presented a number of arguments for
rapid star formation in $\sim$1-2 dynamical timescales, including
scales relevant to star clusters. {\em Tan et al.} (in prep.)
presented arguments for somewhat longer formation timescales and
argued that star formation in clusters is a quasi-equilibrium process.
For example, the age spread of stars in the Orion Nebula Cluster is at
least 2.5\,Myr ({\em Palla and Stahler}, 1999), while the dynamical
time is only $7\times 10^5\:{\rm yr}$ for the cluster as a whole, and
is much shorter in the central region. A relatively long formation
timescale is also consistent with the observed morphologies of
protoclusters in CS molecular lines: {\em Shirley et al.} (2003) found
approximately spherical and centrally concentrated morphologies for a
large fraction of their sources, suggesting they are older than a few
dynamical times. Long formation timescales mean that the observed
central locations of massive stars could be due either to central
formation or mass segregation (or both). A corollary of long formation
timescales is that the level of turbulence in the clump must be
maintained, possibly by protostellar outflows and H{\sc ii} regions
(see \S 2.2, \S 4.2). Studies of the spatial distributions of massive
stars in more embedded, presumably less dynamically evolved, clusters
should help to resolve this issue.

As with spatial segregation, there is also no consensus about whether
there is a temporal segregation in massive star formation from the
surrounding cluster: do massive stars form early, late or
contemporaneously with the other cluster members? Late formation of
massive stars was often proposed, since it was expected that once
massive stars were present, they would rapidly disrupt the remaining
gas with their feedback. However, {\em Tan and McKee} (2001) showed
that the impact of feedback was much reduced in a medium composed of
dense cores, virialized and orbiting supersonically in the clump
potential. {\em Dale et al.} (2005) have carried out the first
simulations of photo-ionizing feedback in clusters and have confirmed
that feedback is significantly reduced in realistic, inhomogeneous
clumps. The observed numbers of UCH{\sc ii} regions ({\em Kurtz et
  al.}, 1994) also suggest that massive star ionizing feedback can be
confined inside $\sim 0.1$\,pc for at least $\sim 10^5\:{\rm yr}$.
{\em Hoogerwerf et al.} (2001) have proposed that four O- and B-stars
were ejected 2.5\,Myr ago from the Orion Nebula Cluster, where massive
star formation is still underway. If true, this would indicate that
massive star formation occurred in both the early and late stages of
cluster formation.

\section{Conclusions and future prospects}
\label{conclusion}

Observational evidence suggests that stars at least up to
30\,M$_{\odot}$ form via an accretion based formation scenario.
Venturing to higher mass objects is an important observational future
task. Theoretically, stars of all masses are capable of forming via
accretion processes but it remains an open question whether Nature
follows that path or whether other processes become more important for
the highest-mass stars. Recent work suggests that the accretion-based
formation scenario in turbulent molecular cloud cores is the more
likely way to build most stars of all masses.

Regarding the evolutionary sequence outlined in the Introduction, the
HMPO and Final-Star stages have been studied extensively in the past,
whereas the earliest stages of massive star formation, i.e., High-Mass
Starless Cores and High-Mass Cores harboring Low/Intermediate-Mass
Protostars, are just beginning to be explored in more detail. The
coming years promise important results for the initial conditions of
massive star formation and the origin of the IMF.

One of the observational challenges of the coming decade is to
identify and study the properties of genuine accretion disks in
high-mass star formation (see also the chapter by {\em Cesaroni et
  al.}). Are massive accretion disks similar to their low-mass
counterparts, or are they massive enough to become self-gravitating
entities? Determining the nature of the broad Radio Recombination
Lines in HCH{\sc ii} regions requires high spatial resolution and high
sensitivity, which both will be provided by the EVLA and ALMA.
Ultimately, we need to understand how outflows are collimated and
driven. Do they originate from the surface of the disk?  What fraction
of the matter that becomes unstable and begins falling toward the
star/disk actually makes it into the star versus being thrown back out
via bipolar outflows? What fraction of the outflow mass is due to gas
entrainment and what fraction is due to recycled infalling gas?  At
what evolutionary stage is the IMF actually determined? Furthermore,
astrochemistry is still poorly investigated, but the advent of large
correlator bandwidths now allows us to investigate the chemical census
in massive star-forming regions regularly in more detail. An important
astrochemical goal is to establish chemical clocks for star-forming
clumps and cores.

The observational capabilities available now and coming online within
the next few years are exciting. To mention a few: The SPITZER
observatory, and especially the SPITZER surveys GLIMPSE and MIPSGAL
will provide an unprecedented census of star-forming regions over
large parts of the Galactic Plane. The so far poorly explored
far-infrared spectrum will be available with the launches of SOFIA and
Herschel. The SMA is currently opening the submm spectral window to
high spatial resolution observations, and ALMA will revolutionize
(sub)mm interferometry and star formation research in many ways. Near-
and mid-infrared interferometry is still in its infancy but early
results from the VLTI are very promising. In addition, many existing
observatories are upgraded to reach new levels of performance (e.g.,
PdBI, EVLA, CARMA).  Combining the advantages of all instruments,
massive star formation research is going to experience tremendous
progress in the coming years.

Theorists face the same challenges as observers in understanding the
formation and evolution of the molecular clouds and clumps that are
the sites of massive star formation, the processes by which individual
and binary massive stars form, the origin of the IMF, the strong
feedback processes associated with massive star formation, and the
interactions that occur in stellar clusters.  Here the primary
progress is likely to come from simulations on increasingly powerful
computers.  By the time of the next Protostars and Planets meeting, it
should be possible to simulate the formation of a cluster of stars in
a turbulent, magnetized medium, to assess the merits of existing
theoretical models, and to point the way toward a deeper understanding
of massive star formation.\\

\textbf{ Acknowledgments.} We wish to thank M.~Krumholz and S.~Lizano
for helpful comments. The research of C.F.M.~is supported in part by
NSF grant AST00-98365. E.C.~acknowledges partial support by NSF grant
978959.  H.B.~acknowledges financial support by the
Emmy-Noether-Program of the Deutsche Forschungsgemeinschaft (DFG,
grant BE2578).

\centerline\textbf{REFERENCES}
\parskip=0pt
{\small
\baselineskip=11pt

\refs Alcolea J., Menten K. M., Moran J. M., and Reid M. J.~(1993)
{\em Lecture Notes in Physics, 412}, 225-228. Springer, Berlin.

\refs Andr{\'e} P., Ward-Thompson D., and Barsony M.~(1993) {\em \apj,
406}, 122-141.

\refs Bacmann A., Andr{\'e} P., Puget J.-L., Abergel A., Bontemps S.,
et al.~(2000) {\em \aap, 361}, 555-580.

\refs Ballesteros-Paredes J., Hartmann L., and Vazquez-Semadeni
E.~(1999) {\em \apj, 527}, 285-297.

\refs Ballesteros-Paredes J., Gazol A., Kim J., Klessen R. S., Jappsen
A.-K., et al.~(2006) {\em \apj, accepted}, astro-ph/0509591.

\refs Bally J.~and Zinnecker H.~(2005) {\em \aj, 129}, 2281-2293.

\refs Behrend R. and Maeder A.~(2001) {\em \aap, 273}, 190-198.

\refs Belloche A., Andr{\'e} P., and Motte F.~(2001) In {\em From Darkness to
  Light}, (T. Montmerle and P. Andr{\'e}, eds.), 313-318. ASP 243, San
Francisco.

\refs Beltr{\'a}n M. T., Cesaroni R., Neri R., Codella C., Furuya
R. S., et al.~(2004) {\em \apj, 601}, L187-L190.

\refs Beltr{\'a}n M. T., Brand J., Cesaroni C., Fontani F., Pezzuto
S., et al.~(2006) {\em \aap, accepted}, astro-ph/0510422.

\refs Benjamin R. A., Churchwell E., Babler B. L., Bania T. M.,
Clemens D. P., et al.~(2003) {\em Publ. Astron. Soc. Pac., 115}, 953-964

\refs Bertoldi F. and McKee C. F.~(1992) {\em \apj, 395}, 140-157.

\refs Beuther H. and Schilke P.~(2004) {\em Science, 303}, 1167-1169.

\refs Beuther H. and Shepherd D.~(2005) In {\em Cores to
Clusters}, (M. S. N. Kumar et al., eds), 105-119. Springer, New York.

\refs Beuther H., Schilke P., Menten K. M., Motte F., Sridharan T. K.,
et al.~(2002a) {\em \apj, 566}, 945-965.

\refs Beuther H., Schilke P., Sridharan T. K., Menten K. M., Walmsley
C. M., et al.~(2002b) {\em \aap, 383}, 892-904.

\refs Beuther H., Walsh A., Schilke P., Sridharan T. K., Menten K. M.,
et al.~(2002c) {\em \aap, 390}, 289-298.

\refs Beuther H., Zhang Q., Greenhill L. J., Reid M. J., Wilner D., et
al.~(2004) {\em \apj, 616}, L31-L34.

\refs Beuther H., Zhang Q., Greenhill L. J., Reid M. J., Wilner D., et
al.~(2005a) {\em \apj, 632}, 355-370.

\refs Beuther H., Zhang Q., Sridharan T. K., and Chen Y.~(2005b)
{\em \apj, 628}, 800-810.

\refs Beuther H., Sridharan T. K., and Saito M.~(2005c) {\em \apj,
634}, L185-L188.

\refs Beuther H., Zhang Q., Reid M. J., Hunter T. R., Gurwell M., et
al.~(2006) {\em \apj, 636}, 323-331.

\refs Birkmann S. M., Krause O., and Lemke D.~(2006) {\em \apj, 637},
380-383.
 
\refs Blake G. A., Mundy L. G., Carlstrom J. E., Padin S., Scott
S. L., et al.~(1996) {\em \apj, 472}, L49-L52.

\refs Blake G. A., Sutton E. C., Masson C. R., and Phillips
T. G.~(1987) {\em \apj, 315}, 621-645.

\refs Bonnell I. A. and Bate M. R.~(2002) {\em \mnras, 336}, 659-669.

\refs Bonnell I. A. and Bate M. R.~(2005) {\em \mnras, 362}, 915-920.

\refs Bonnell I. A. and Davies M. B.~(1998) {\em \mnras, 295}, 691-698.

\refs Bonnell I. A., Bate M. R., and Zinnecker H.~(1998) {\em
\mnras, 298}, 93-102.

\refs Bonnell I. A., Bate M. R., Clarke C. J., and Pringle
J. E.~(2001a) {\em \mnras, 323}, 785-794.

\refs Bonnell I. A., Clarke C. J., Bate M. R., and Pringle J. E.
(2001b) {\em \mnras, 324}, 573-579..

\refs Bonnell I. A., Vine S. G., and Bate M. R.~(2004) {\em \mnras,
349}, 735-741.

\refs Bonnell I. A., Dobbs C. L., Robitaille T. P., and Pringle
J. E.~(2006) {\em \mnras, 365}, 37-45.

\refs Bourke T. L., Myers P. C., Robinson G., and Hyland A. R.~(2001)
{\em \apj, 554}, 916-932.

\refs Bromm V. and Loeb A.~(2004) In {\em The Dense Interstellar Medium
in Galaxies}, (S. Pfalzner et al., eds.), 3-10. Springer, Berlin.

\refs Bronfman L., Nyman L.-A., and May J.~(1996) {\em \apjs,
115}, 81-95.

\refs Carey S. J., Feldman P. A., Redman R. O., Egan M. P., MacLeod,
J. M., et al.~(2000) {\em \apj, 543}, L157-L161.

\refs Carral P., Kurtz S. E., Rodr{\'{\i}}guez L. F., de Pree C., and
Hofner P.~(1997) {\em \apj, 486}, L103-L106

\refs Cesaroni R.~(2005) In {\em \apss, 295}, 5-17.

\refs Cesaroni R., Felli M., Testi L., Walmsley C. M., and Olmi L.~(1997)
{\em \aap, 325}, 725-744.

\refs Cesaroni R., Felli M., Jenness T., Neri R., Olmi L., et
al.~(1999) {\em \aap, 345}, 949-964.

\refs Cesaroni R., Neri R., Olmi L., Testi L., Walmsley C. M., et
al.~(2005a) {\em \aap, 434}, 1039-1054.

\refs Cesaroni R., Felli M., Churchwell E., and Walmsley C. M.~(2005b)
{\em eds. IAU227: Massive Star Birth: A Crossroad to Astrophysics}.
Cambridge Univ., Cambridge.

\refs Chakrabarti S. and McKee C.~(2005) {\apj, 631}, 792-808.

\refs Charnley S. B.~(1997) {\em \apj, 481}, 396-405.

\refs Chieze J.-P.~(1987) {\em \aap, 171}, 225-232.

\refs Cho J. and Lazarian A.~(2003) {\em \mnras, 345}, 325-339.

\refs Churchwell E.~(2002) {\em Ann. Rev. Astron. Astrophys., 40}, 27-62.

\refs Churchwell E., Walmsley C. M., and Cesaroni R.~(1990) {\em
\aaps, 83}, 119-144.

\refs Clark P. C., Bonnell I. A., Zinnecker H., and Bate
M. R.~(2005) {\em \mnras, 359}, 809-818.

\refs Codella C., Lorenzani A., Gallego A. T., Cesaroni R., and
Moscadelli L.~(2004) {\em \aap, 417}, 615-624.

\refs Corbelli B, Palla F., and Zinnecker H.~(2005) {\em eds. The
Initial Mass Function 50 years later}. Springer, Dordrecht.

\refs Crutcher R. M.~(1999) {\em \apj, 520}, 706-713.

\refs Dale J. E., Bonnell I. A., Clarke C. J., and Bate M. R.~
(2005) {\em \mnras, 358}, 291-304.

\refs Dame T. M., Elmegreen B. G., Cohen R. S., and Thaddeus
P.~(1986) {\em \apj, 305}, 892-908.

\refs De Buizer J. M.~(2003) {\em \mnras, 341}, 277-298.

\refs De Buizer J. M., Watson A. M., Radomski J. T., Pina R. K., and
Telesco C. M.~(2002) {\em \apj, 564}, L101-L104.

\refs De Buizer J. M., Osorio M., and Calvet N.~(2005) {\apj,
635}, 452-465.

\refs De Pree C. G., Wilner D. J., Goss W. M., Welch W. J., and
McGrath E.~(2000) {\em \apj, 540}, 308-315.

\refs De Wit W. J., Testi L., Palla F., and Zinnecker H.~(2005)
{\em \aap, 437}, 247-255.

\refs Dobbs C. L, Bonnell I. A., and Clark P. C.~(2005) {\em \mnras, 
360}, 2-8.

\refs Doty S. D., van Dishoeck E. F., van der Tak F. F. S., and Boonman
A. M.~(2002) {\em \aap, 389}, 446-463.

\refs Draine  B. T.~(2003) {\em Ann. Rev. Astron. Astrophys., 41}, 241-289.

\refs Edgar R. and Clarke C. J.~(2004) {\em \mnras, 349}, 678-686.

\refs Egan M. P., Shipman R. F., Price S. D., Carey S. J., Clark
F. O., et al.~(1998) {\em \apj, 494}, L199-L202.

\refs Ellingsen S. P.~(2006) {\em \apj, accepted}, astro-ph/0510218.

\refs Elmegreen B. G.~(1989) {\em \apj, 338}, 178-196.

\refs Elmegreen B. G.~(2000) {\em \apj, 530}, 277-281.

\refs Elmegreen B. G. and Scalo J.~(2004) {\em
Ann. Rev. Astron. Astrophys., 42}, 211-273.

\refs Enoch M. L., Young K. E., Glenn J., Evans N. J., Golwala S., et
al.~(2006) {\em \apj, accepted}, astro-ph/0510202.

\refs Evans N. J.~(1999) {\em Ann. Rev. Astron. Astrophys., 37}, 311-362.

\refs Evans N. J., Shirley Y. L., Mueller K. E., and Knez C.~(2002)
In {\em Hot Star Workshop III: The Earliest Stages of Massive Star
Birth} (P. A. Crowther, ed.), 17-31. ASP 267, Michigan.

\refs Faison M., Churchwell E., Hofner P., Hackwell J., Lynch D. K.,
et al.~(1998) {\em \apj, 500}, 280-290.

\refs Fa{\'u}ndez S., Bronfman L., Garay G., Chini R., Nyman, L.-A.,
et al.~(2004) {\em \aap, 426}, 97-103.

\refs Fendt C. and Cameljic M.~(2002) {\em \aap,, 395}, 1045-1060.

\refs Figer D. F.~(2005) {\em Nature, 34}, 192-194.

\refs Garay G., Faundez S., Mardones D., Bronfman L., Chini R., et
al.~(2004) {\em \apj, 610}, 313-319.

\refs Gaume R. A., Claussen M. J., de Pree C. G., Goss W. M., and
Mehringer D. M.~(1995) {\em \apj, 449}, 663-673.

\refs Goddi C., Moscadelli L., Alef W., Tarchi A., Brand J., et
al.~(2005) {\em \aap, 432}, 161-173.

\refs Hartmann L., Ballesteros-Paredes J., and Bergin E.~(2001) {\em
\apj, 562}, 852-868.

\refs Hatchell J. and van der Tak F. F.S. ~(2003) {\em \aap, 409},
589-598.

\refs Heitsch F., Burkert A., Hartmann L. W., Slyz A. D., and Devriendt
J. E. G.~(2005) {\em \apj, 633}, L113-l116.

\refs Hill T., Burton M., Minier V., Thompson M. A., Walsh A. J., et
al.~(2005) {\em \mnras, 363}, 405-451.

\refs Hofner P. and Churchwell E.~(1996) {\em \aaps, 120}, 283-299.

\refs Hofner P., Kurtz S., Churchwell E., Walmsley C. M., and Cesaroni
R.~(1996) {\em \apj, 460}, 359-371.

\refs Hollenbach D., Johnstone D., Lizano S., and Shu F.~(1994)
{\em \apj, 428}, 654-669.

\refs Hoogerwerf R., de Bruijne J. H. J., and de Zeeuw P. T.~(2001)
{\em \aap, 365}, 49-77.

\refs Hunter T. R., Churchwell E., Watson C., Cox P., Benford D. J.,
et al.~(2000) {\em \aj, 119}, 2711-2727.

\refs Ignace R. and Churchwell E.~(2004) {\em \apj, 610}, 351-360.

\refs Indebetouw R., Whitney B. A., Johnson K. E., and Wood
K.~(2006) {\em \apj, 636}, 362.

\refs Jappsen A.-K. and Klessen R. S.~(2004) {\aap, 423}, 1-12.

\refs Jijina J. and Adams F. C.~(1996) {\em \apj,  462}, 874-884.

\refs Johnstone D., Fich M., Mitchell G. F., and Moriarty-Schieven
G.~(2001) {\em \apj, 559}, 307-317.

\refs Kahn F. D.~(1974) {\em \aap, 37}, 149-162.

\refs Keto E.~(2002) {\em \apj, 580}, 980-986.

\refs Keto E.~(2003) {\em \apj, 599}, 1196-1206.

\refs Keto E. and Wood K.~(2006) {\em \apj, accepted},
astro-ph/0510176.

\refs Kim K.-T. and Kurtz S. E.~(2006) {\em \apj, accepted},
astro-ph/0601532

\refs Klein R., Posselt B., Schreyer K., Forbrich J., and Henning
T.~(2005) {\em \apjs, 161}, 361-393.

\refs Kramer C., Stutzki J., R\"ohrig R., and Corneliussen U.~(1998)
{\em \aap, 329}, 249-264.

\refs Krumholz M. R., McKee C. F., and Klein R. I.~(2005a) {\em \apj, 
618}, L33-L36.

\refs Krumholz M. R., Klein R. I., McKee C. F.~(2005b) In {\em Massive
  Star Birth: A Crossroads of Astro-Physics, IAU 227} (Cesaroni et
al., eds.), 231-236. Cambridge Univ., Cambridge.

\refs Krumholz M. R., McKee C. F., and Klein R. I.~(2005c)
{\em Nature, 438}, 332-334.

\refs Krumholz M. R., McKee C. F., and Klein R. I.~(2006)
{\em \apj}, in press~(astro-ph/0510410).

\refs Kurtz S., Churchwell E., and Wood D. O. S.~(1994) {\em \aaps, 91},
659-712.

\refs Kurtz S.~(2002) In {\em Hot Star Workshop III: The Earliest
  Stages of Massive Star Birth} (P. A. Crowther, ed.), 81-94. ASP 267,
Michigan.

\refs Kurtz S.~(2005) In {\em Massive Star Birth: A Crossroads of
  Astrophysics, IAU 227} (Cesaroni et al., eds.), 111-119. Cambridge
Univ., Cambridge.

\refs Kylafis N. D. and Pavlakis K. G.~(1999) In {\em The Origins of
  Stars and Planetary Systems} (C. J. Lada and N. D. Kylafis, eds.),
553-575. Kluwer, Dordrecht.

\refs Lada C. L. and Wilking B. A.~(1984) {\em \apj, 287}, 610-621.

\refs Larson R.~(1981) {\em MNRAS, 194}, 809-826.

\refs Linz H., Stecklum B., Henning T., Hofner P., and Brandl
B.~(2005) {\em \aap, 429}, 903-921

\refs Mac Low M.~(2004) In {\em The Dense Interstellar Medium
in Galaxies}, (S. Pfalzner et al., eds.), 379-386. Springer, Berlin.

\refs Mac Low M. and Klessen R. S.~(2004) {\em Rev. Mod.
Phys., 76}, 125-194.

\refs Matt S. and Pudritz R. E.~(2005) {\em \mnras, 356}, 167-182.

\refs Matzner C. D. and McKee C. F.~(2000) {\em \apj, 545}, 364-378.

\refs Matzner C. D.~(2002) {\em \apj, 566}, 302-314.

\refs McKee C. F.~(1989) {\em \apj, 345}, 782-801.

\refs McKee C. F.~(1999) In {\em The Origins of Stars and Planetary
  Systems} (C. J. Lada and N. D. Kylafis, eds.), 29-67.  Kluwer,
Dordrecht.

\refs McKee C. F. and Tan J. C.~(2002) {\em Nature, 416}, 59-61.

\refs McKee C. F. and Tan J. C.~(2003) {\em \apj, 585}, 850-871.

\refs McKee C. F. and Chakrabarti S.~(2005) In {\em Massive Star
  Birth: A Crossroads of Astrophysics, IAU 227} (Cesaroni et al.,
eds.), 276-281. Cambridge Univ., Cambridge.

\refs McLaughlin D. E. and Pudritz R. E.~(1997) {\em \apj, 476}, 750-765.

\refs Megeath S. T., Wilson T. L., and Corbin M. R.~(2005) {\em \apj,
622}, 141-144.

\refs Meynet G. and Maeder A.~(2005) In {\em The Nature and Evolution
of Disks Around Hot Stars} (Ignace R. and Gayley K. G., eds.),
15-26. ASP Conf. Series 337, Tennessee.

\refs Molinari S., Brand J., Cesaroni R., and Palla F.~(1996)
{\em \aap, 308}, 573-587.

\refs Molinari S., Brand J., Cesaroni R., Palla F., and Palumbo
G. G. C.~(1998) {\em \aap, 336}, 339-351.

\refs Motte F., Andr{\'e} P., and Neri R.~(1998) {\em \aap, 336}, 150-172.

\refs Motte F. and Andr{\'e} P.~(2001) {\em \aap, 365}, 440-464.

\refs Motte F., Bontemps S., Schilke P., Lis D. C., Schneider N., et
al.~(2005) In {\em Massive Star Birth: A Crossroads of Astrophysics}
(Cesaroni et al., eds.), 151-156. Cambridge Univ., Cambridge.

\refs Mueller K. E., Shirley Y. L., Evans N. J., and Jacobson
H. R.~(2002) {\em \apjs, 143}, 469-497.

\refs Myers P. C. and Goodman A. A.~(1988) {\em \apj, 326}, L27-L30.

\refs Nakano T.~(1989) {\em \apj, 345}, 464-471.

\refs Nomura H. and Millar T. J.~(2004) {\em \aap, 414}, 409-423.

\refs Norberg P. and Maeder A.~(2000) {\em \aap, 359}, 1025-1034.

\refs Norman C. and Silk J.~(1980) {\em \apj, 238}, 158-174.

\refs Norris R. P., Byleveld S. E., Diamond P. J., Ellingsen S. P.,
Ferris R. H., et al.~(1998) {\em \apj, 508}, 275-285.

\refs Omukai K. and Inutsuka S.~(2002) {\em \mnras, 332}, 59-64.

\refs Onishi T., Mizuno A., Kawamura A., Ogawa H., and Fukui 
Y.~(1996) {\em \apj, 465}, 815-824.

\refs Ormel C. W., Shipman R. F., Ossenkopf V., and Helmich F. P.~(2005)
{\em \aap, 439}, 613-625.

\refs Osorio M., Lizano S., and D'Alessio P.~(1999) {\em \apj,
525}, 808-820.

\refs Ostriker E. C.~(1999) {\em \apj, 513}, 252-258.

\refs Padoan P. and Nordlund A.~(2002) {\em \apj, 576}, 870-879.

\refs Palla F. and Stahler S. W.~(1999) {\em \apj, 525}, 722-783.

\refs Peretto N., Andr{\'e} P., and Belloche A.~(2006) {\em \aap, in press},
astro-ph/0508619.

\refs Pestalozzi M. R., Elitzur M., Conway J. E., and Booth
R. S.~(2004) {\em \apj, 603}, L113-L116.

\refs Pestalozzi M. R., Minier V., and Booth R. S.~(2005) {\em \aap,
432}, 737-742.

\refs Pillai T., Wyrowski F., Menten K. M., and Kr\"ugel E.~(2006) {\em
\aap, in press}, astro-ph/0510622.

\refs Plume R., Jaffe D. T., and Evans N. J.~(1992) {\em \apjs, 78}, 505-515.

\refs Plume R., Jaffe D. T., Evans N. J, Martin-Pintado J., and
Gomez-Gonzales J.~(1997) {\em \apj, 476}, 730-749.

\refs Portegies Zwart S. F., Baumgardt H., Hut P., Makino J.,
and McMillan S. L. W.~(2004) {\em Nature, 428}, 724-726.

\refs Quillen A. C., Thorndike S. L., Cunningham A., Frank A.,
Gutermuth R. A., et al.~(2005) {\em \apj, 632}, 941-955.

\refs Rathborne J. M., Jackson J. M., Chambers E. T., Simon R.,
Shipman R., et al.~(2005) {\em \apj, 630}, L181-L184.

\refs Reid M. A. and Wilson C. D.~(2005) {\em \apj, 625}, 891-905.

\refs Richer J. S., Shepherd D. S., Cabrit S., Bachiller R., and
Churchwell E.~(2000) In {\em Protostars and Planets IV} (V. Mannings
et al., eds.), 867-894.  Univ. of Arizona, Tucson.

\refs Rudolph A., Welch W. J., Palmer P., and Dubrulle B.~(1990) {\em
\apj, 363}, 528-546.

\refs Salpeter E. E.~(1955) {\em \apj, 121}, 161-167.

\refs Sandell G, Wright M., and Forster J. R.~(2003) {\em \apj,
590}, L45-L48.

\refs Sanders D. B., Scoville N. Z., Tilanus R. P. J., Wang Z.,
and Zhou S.~(1993) In {\em Back to the Galaxy}, (S. S. Holt
and F. Verter, eds.), 311-314. AIP 278, New York.

\refs Scalo J. and Elmegreen B. G.~(2004) {\em
Ann. Rev. Astron. Astrophys., 42}, 275-316.

\refs Schilke P., Groesbeck T. D., Blake G. A., and Phillips
T. G.~(1997) {\em \apj, 108}, 301-337.

\refs Schilke P., Benford D. J., Hunter T. R., Lis D. C., and
Phillips T. G.~(2001) {\em \apjs, 132}, 281-364.

\refs Schmeja S. and Klessen R. S.~(2004) {\em \aap, 419}, 405-417.

\refs Sewilo M., Churchwell E., Kurtz S., Goss W. M., and Hofner
P.~(2004) {\em \apj, 605}, 285-299.

\refs Shepherd D. S. and Churchwell E.~(1996) {\em \apj, 457}, 267-276.

\refs Shepherd D. S., Testi L., and Stark D. P.~(2003) {\em \apj,
584}, 882-894.

\refs Shirley Y. L., Evans N. L., Young K. E., Knez C., and Jaffe
D. T.~(2003) {\em \apjs, 149}, 375-403.

\refs Solomon P. M., Rivolo A. R., Barrett J. W., and Yahil
A.~(1987) {\em \apj, 319}, 730-741.

\refs Spitzer L. Jr.~(1978) {\em Physical Processes in the Interstellar
Medium}, Wiley, New York.

\refs Sridharan T. K., Beuther H., Schilke P., Menten K. M.,
and Wyrowski F.~(2002) {\em \apj, 566}, 931-944.

\refs Sridharan T. K., Beuther H., Saito M., Wyrowski F., and Schilke
P.~(2005) {\em \apj, 634}, L57-L60.

\refs Stahler S. W., Shu F. H., and Taam R. E. 1980, \apj, 241, 637-654.

\refs Stahler S. W., Palla F., and Ho P. T. P.~(2000) In {\em
  Protostars and Planets IV} (V. Mannings et al., eds.), 327-351.
Univ. of Arizona, Tucson.

\refs Tan J. C. and McKee C. F.~(2001) In {\em Starburst Galaxies:
  Near and Far} (L. Tacconi et al., eds.), 188-196. Springer,
Heidelberg.

\refs Tan J. C. and McKee C. F.~(2003) In {\em Star Formation at High
  Angular Resolution, IAU 221}, Poster 274, astro-ph/0309139.

\refs Tan J. C.~(2006) In {\em The Young Local Universe}, astro-ph/0407093.

\refs Torrelles J. M., G{\'o}mez J. F., Rodr{\'{\i}}guez L. F., Curiel S.,
Anglada G., et al.~(1998) {\em \apj, 505}, 756-765.

\refs Torrelles J. M., Patel N. A., Anglada G., G{\'o}mez J. F., Ho
P. T. P., et al.~(2003) {\em \apj, 598}, L115-L118.

\refs Ulrich R. K.~(1976) {\em \apj, 210}, 377-391.

\refs van der Tak F. F.S. and Menten K. M.~(2005) {\em \aap, 437}, 947-956.

\refs van Dishoeck E. F. and Blake G. A.~(1998) {\em
Ann. Rev. Astron. Astrophys., 36}, 317-368.

\refs Viti S., Collings M. P., Dever J. W., McCoustra M. R. S.,
and Williams D. A.~(2004) {\em \mnras, 354}, 1141-1145.

\refs Walmsley C. M.~(1995) {\em Rev. Mex. Astron. 
Astrof., 1}, 137-148.

\refs Walsh A. J., Burton M. G., Hyland A. R., and Robinson
G.~(1998) {\em \mnras, 301}, 640-698.

\refs Walsh A. J., Macdonald G. H., Alvey N. D. S., Burton M. G., and
Lee J.-K.~(2003) {\em \aap, 410}, 597-610.

\refs Weidner C. and Kroupa P.~(2004) {\em \mnras, 348}, 187-191.

\refs Whitney B. A., Robitaille T. P., Indebetouw R., Wood K.,
Bjorkmann J. E., et al.~(2005) In {\em Massive Star Birth: A
Crossroads of Astrophysics, IAU 227} (Cesaroni et al., eds.),
206-215. Cambridge Univ., Cambridge.

\refs Williams J. P., Blitz L., and McKee C. F.~(2000) In {\em
  Protostars and Planets IV} (V. Mannings et al., eds.), 97-120.
Univ. of Arizona, Tucson.

\refs Williams J. P. and Garland C. A.~(2002) {\em \apj, 568},
259-266.

\refs Williams S. J., Fuller G. A., and Sridharan T. K.~(2004) {\em
\aap, 417}, 115-133.

\refs Williams S. J., Fuller G. A., and Sridharan T. K.~(2004) {\em
\aap, 434}, 257-274.

\refs Wilner D. J., Reid M. J., and Menten K. M.~(1999) {\em \apj,
513}, 775-779.

\refs Wilson T. L., Gaume R. A., Gensheimer P., and Johanston
K. T.~(2000) {\em \apj, 538}, 665-674.

\refs Wolfire M. G. and Cassinelli J. P.~(1987) {\em \apj, 319},
850-867.

\refs Wood D. O. S. and Churchwell E.~(1989a) {\em \apj, 340}, 265-272.

\refs Wood D. O. S. and Churchwell E.~(1989b) {\em \apjs, 69}, 831-895.

\refs Wright M. C. H., Plambeck R. L., and Wilner D. J.~(1996) {\em
\apj, 469}, 216-237.

\refs Wyrowski F., Hofner P., Schilke P., Walmsley C. M., Wilner
D. J., et al.~(1997) {\em \aap, 320}, L17-L20.

\refs Wyrowski F., Schilke P., Walmsley C. M., and Menten
K. M.~(1999) {\em \apj, 514}, L43-L46.

\refs Yonekura Y., Asayama S., Kimura K., Ogawa H., Kanai Y., et
al.~(2005) {\em \apj, 634}, 476-494.

\refs Yorke H. W. and Sonnhalter C.~(2002) {\em \apj, 569}, 846-862.

\refs Zhang Q., Hunter T. R., and Sridharan T. K.~(1998) {\em \apj,
505}, L151-L154.

\refs Zhang Q., Hunter T. R., Sridharan T. K., and Ho P. T. P.~(2002)
{\em \apj, 566}, 982-992.

\refs Zhang Q., Hunter T. R., Brand J., Sridharan T. K., Cesaroni R.,
et al.~(2005) {\em \apj, 625}, 864-882.  }

\end{document}